# One Hundred Years of the Cosmological Constant: from 'Superfluous Stunt' to Dark Energy

Cormac O'Raifeartaigh,[a] Michael O'Keeffe,[a] Werner Nahm[b] and Simon Mitton[c]

[a]*School of Science and Computing, Waterford Institute of Technology, Cork Road, Waterford, Ireland*
[b]*School of Theoretical Physics, Dublin Institute for Advanced Studies, 10 Burlington Road, Dublin 2, Ireland*
[c]*St Edmund's College, University of Cambridge, Cambridge CB3 0BN, United Kingdom*
Author for correspondence: coraifeartaigh@wit.ie

## Abstract

We present a centennial review of the history of the term known as the cosmological constant. First introduced to the general theory of relativity by Einstein in 1917 in order to describe a universe that was assumed to be static, the term fell from favour in the wake of the discovery of the expanding universe, only to make a dramatic return in recent times. We consider historical and philosophical aspects of the cosmological constant over four main epochs; (i) the use of the term in static cosmologies (both Newtonian and relativistic): (ii) the marginalization of the term following the discovery of cosmic expansion: (iii) the use of the term to address specific cosmic puzzles such as the timespan of expansion, the formation of galaxies and the redshifts of the quasars: (iv) the re-emergence of the term in today's Λ-CDM cosmology. We find that the cosmological constant was never truly banished from theoretical models of the universe, but was marginalized by astronomers for reasons of convenience. We also find that the return of the term to the forefront of modern cosmology did not occur as an abrupt paradigm shift due to one particular set of observations, but as the result of a number of empirical advances such as the measurement of present cosmic expansion using the Hubble Space Telescope, the measurement of past expansion using type SN 1a supernovae as standard candles, and the measurement of perturbations in the cosmic microwave background by balloon and satellite. We give a brief overview of contemporary interpretations of the physics underlying the cosmic constant and conclude with a synopsis of the famous *cosmological constant problem*.

# 1. Introduction

First introduced one hundred years ago, the mathematical entity known as the cosmological constant plays a central role in modern cosmology. However, the term has had a chequered career over the century. Initially added by Einstein to the field equations of general relativity in order to allow a consistent model of a universe that was assumed to be static (Einstein 1917a), it was gradually sidelined following the emergence of evidence for an expanding universe in the 1930s. Yet the term was invoked on several occasions over the next few decades as a tool to address diverse empirical puzzles such as the timespan of cosmic expansion, the formation of galaxies and the redshifts of quasars. The term also found application in alternate cosmologies such as the steady-state model of the universe, and more recently, in the theory of cosmic inflation. Later still, it was invoked in the 1990s to address puzzles concerning anisotropies in the cosmic microwave background (CMB) and problems concerning the cold dark matter (CDM) model of structure formation. With the discovery of an acceleration in cosmic expansion, sophisticated measurements of cosmic geometry, and new advances in the study of galaxy formation and galaxy clustering, the term returned to the centre stage of modern cosmology in the early years of the 21$^{st}$ century.

The aim of this article is to provide a comprehensive historical review of the story of the cosmological constant. Our review will draw on primary materials made available online by the Einstein Papers Project[1] and on original papers in cosmology, astronomy and astrophysics. We will also draw on classic reviews of the cosmological constant such as (Zel'dovich 1968; McCrea 1971; Weinberg 1989; Carroll, Press and Turner 1992) and more recent 'post-dark energy' discussions such as (Straumann 1999; Earman 2001; Rugh and Zinkernagel 2002; Carroll 2001; Peebles and Ratra 2003; Ellis 2003; Turner and Huterer 2007). While some material is technical by nature, an effort will be made to present a coherent narrative that is comprehensible to physicists and historians alike.

Our article begins with a brief description of the use of a term similar to the cosmological constant in Newtonian cosmology. In section 3, we recall Einstein's use of the term in his 1917 'static' model of the universe, the alternate model proposed by Willem de Sitter and the debate between the two.[2] The use of the term in the 1920s in the cosmologies of Alexander Friedman and Georges Lemaître is recalled in section 4; this section also includes a brief description of early ideas concerning the zero-point energy of the vacuum. In section 5, different attitudes to the cosmological constant following the discovery of cosmic expansion are described. In sections 6 and 7, we recall the use of the term to address various issues in theoretical and observational cosmology in the years 1940-1970, while the relevance of the term for inflationary cosmology is described in section 8. The gradual re-evaluation of the cosmological constant in the 1990s is discussed in section 9, and the emergence of the term as a centrepiece of today's Λ-CDM model of the universe is described in section 10. We give a brief review of contemporary interpretations of the physics underlying the cosmological constant in section 11 and conclude with a brief summary in section 12.

---

[1] The Einstein Papers Project is an invaluable online historical archive of primary sources provided by Princeton University Press in conjunction with the California Institute of Technology and the Hebrew University of Jerusalem. http://einsteinpapers.press.princeton.edu/.

## 2. A cosmological constant in Newtonian cosmology

In modern usage, the term 'cosmological constant' refers to a mathematical term added to the field equations of general relativity in order to give a consistent model of the universe. However, it is worth recalling that a similar entity was employed in Newtonian cosmology, long before the advent of the general theory of relativity.

Towards the end of the 19th century, the eminent German astronomer Hugo von Seeliger noted that, assuming a cosmos of infinite spatial extent[3] and a non-zero, uniform density of matter on the largest scales, the Newtonian gravitational force at any point is indeterminate (Seeliger 1895, 1896).[4] The problem had been noted previously, but Seeliger was the first to address the paradox in a systematic manner (North 1965, pp 16-17; Norton 1999). As he later remarked: *"About two years ago, I drew attention to certain difficulties arising out of the attempt to extend Newton's law of gravitation to infinite space. The considerations then adduced showed the necessity of choosing between two hypotheses, viz: 1. The sum total of the masses of the universe is infinitely great, in which case Newton's law cannot be regarded as a mathematically exact expression for the attractive forces in operation. 2. The Newtonian law is rigourously exact, in which case the infinite spaces of the*

---

[2] This section will draw on our recent historical review of Einstein's 1917 model of the universe (O'Raifeartaigh et al. 2017).

[3] This assumption was necessary in order to avoid the problem of gravitational collapse, as first pointed out by Richard Bentley (Ray 1990; Norton 1999; Kragh 2007 pp 72-74).

[4] These papers have not been translated into English, but Seeliger summarized his work on the problem in English two years later (Seeliger 1898a, 1898b).

*universe cannot be filled with matter of finite density. Inasmuch as I am wholly unable to find adequate reasons for the second of these assumptions, I have, in another place, decided in favour of the first"* (Seeliger 1898a).

Seeliger's solution was to modify the familar Newtonian force with an exponential decay factor that would become significant only at extremely large distances. Expressing his argument in terms of the gravitational potential

$$\Phi(r) = G \int \frac{\rho}{r} dV \qquad (1)$$

where G is Newton's gravitational constant and $\rho$ is the mean density of matter in a volume element of space $dV$, Seeliger noted first that the Newtonian potential would not be defined as the distance $r$ tended to infinity. His suggestion was to redefine the potential according to

$$\Phi(r) = G \int \frac{\rho}{r} e^{-\sqrt{\lambda}r} dV \qquad (2)$$

where $\lambda$ was a decay constant sufficiently small to make the modification significant only at extremely large distances. In terms of classical mechanics, this modification amounted to replacing the well-known Poisson equation

$$\nabla^2 \Phi = 4\pi G\rho \qquad (3)$$

with the relation

$$\nabla^2 \Phi - \lambda \Phi = 4\pi G\rho \qquad (4)$$

where $\nabla^2$ is the Laplacian operator. A similar modification of the Newtonian potential - albeit in a slightly different context - had earlier been suggested by physicists such as Pierre-Simon Laplace (Laplace 1846) and Carl Neumann (Neumann 1896 pp 373-379).[5]

Thus, we note that the concept of a cosmological constant - in the sense of a constant decay term added to standard gravitation theory in order to give a consistent model of the universe - was a feature of theoretical cosmology before the advent of relativistic models of the universe. It is also interesting to note that Seeliger suggested that an estimate of λ might be obtained from astronomical measurement of anomalies in the orbits of the planets (Seeliger 1895), although this proposal was not successful (Seeliger 1896; Norton 1999).

---

[5] See (Ray 1990) or (Norton 1999).

## 3. The cosmological constant in the 1910s: the cosmologies of Einstein and de Sitter

In November 1915, Einstein published the field equations of the general theory of relativity. This was a set of covariant equations that related the geometry of a region of space-time to the distribution of matter/energy within it, according to

$$G_{\mu\nu} = -\kappa \left( T_{\mu\nu} - \frac{1}{2} g_{\mu\nu} T \right) \tag{5}$$

Here, $G_{\mu\nu}$ is a four-dimensional tensor representing the curvature of space-time (known as the Ricci curvature tensor), $T_{\mu\nu}$ is a four-dimensional tensor representing energy and momentum, $T$ is a scalar, and $\kappa$ is the Einstein constant $8\pi G/c^2$ (Einstein 1915a). A few years later, it was realised that the general field equations could be written in simpler form as

$$G_{\mu\nu} - \frac{1}{2} g_{\mu\nu} G = -\kappa\, T_{\mu\nu} \tag{6}$$

where $G\ (=\kappa T)$ is a scalar known as the Ricci curvature scalar.[6] Thus, in the general theory of relativity, the flat 'Minkowskian' space-time interval of special relativity

$$ds^2 = -dx^2 - dy^2 - dz^2 + c^2 dt^2 \tag{7}$$

is replaced by a space-time interval of curved geometry given by

$$ds^2 = \sum_{\mu,\nu=0}^{3} g_{\mu\nu} dx^\mu dx^\nu \tag{8}$$

---

[6] We employ the nomenclature used by Einstein in the years 1915-1917. Nowadays, the Ricci curvature tensor and Ricci scalar are denoted by $R_{\mu\nu}$ and $R$ respectively.

where the elements $g_{\mu\nu}$ of the space-time tensor are determined by the distribution and flux of matter/energy. In his seminal papers of 1915 and 1916, Einstein noted that general relativity passed an important empirical test; as well as furnishing a description of the orbits of the planets that was compatible with astronomical observation, the new theory accounted for a well-known anomaly for the case of the motion of Mercury (Einstein 1915b, 1916).

Another important test for the general theory of relativity was the issue of a consistent cosmology, i.e., whether the theory could give a consistent description of the universe as a whole. As Einstein later remarked to the Dutch astronomer Willem de Sitter: *"For me, though, it was a burning question whether the relativity concept can be followed through to the finish, or whether it leads to contradictions. I am satisfied now that I was able to think the idea through to completion without encountering contradictions"* (Einstein 1917b). Indeed, it is clear from Einstein's correspondence of 1916 that cosmic considerations were a major preoccupation for him in the immediate aftermath of the discovery of the covariant field equations (Realdi and Peruzzi 2009; O'Raifeartaigh et al. 2017).

Einstein's starting point for his 1917 model of the cosmos was the assumption of a static, non-zero distribution of matter, and therefore a static metric of positive spatial curvature. As he remarked in the paper: *"The most important fact that we draw from experience as to the distribution of matter is that the relative velocities of the stars are very small as compared with the velocity of light. So I think that for the present we may base our reasoning upon the following approximate assumption. There is a system of reference relatively to which matter may be looked upon as being permanently at rest"* (Einstein 1917a). This assumption was reasonable at the time; many years were to elapse before the discovery of a linear relation between the recession of the distant galaxies and their distance (Hubble 1929), the first evidence for a non-static universe.

A second assumption was that of a uniform distribution of matter (Einstein 1917a). This assumption implied a universe that was both isotropic and homogeneous on the largest scales and was later named the 'Cosmological Principle' (Milne 1935 p24). One reason for the principle was simplicity, as it greatly simplified the business of solving the field equations. A deeper reason was that the principle chimed with the spirit of relativity and with a Copernican approach to cosmology. After all, to assume otherwise was to posit a universe in which all viewpoints were not equivalent; indeed, it's worth noting that the Cosmological Principle was originally named "the extended principle of relativity" (Milne 1933).

Two further assumptions were necessary for Einstein's 1917 model of the cosmos. First, he was obliged to postulate a closed spatial geometry for the cosmos in order to render his model consistent with his views on the relativity of inertia.[7] Second, he found it necessary to add a new term to the general field equations in order to avoid an inconsistent solution. Thus, only 15 months after the publication of the field equations of general relativity (5), Einstein proposed a modification of the equations according to

$$G_{\mu\nu} - \lambda g_{\mu\nu} = -\kappa \left(T_{\mu\nu} - \tfrac{1}{2} g_{\mu\nu} T\right) \qquad (9)$$

Here the $g_{\mu\nu}$ represent the familiar components of the spacetime metric and $\lambda$ is a universal constant that became known as the cosmological constant (Einstein 1917a). Certainly, relativity allowed the term. As Einstein pointed out in his 1917 paper: *"The system of equations… allows a readily suggested extension which is compatible with the relativity postulate….for on the left-hand side of the field equation… we may add the fundamental tensor $g_{\mu\nu}$, multiplied by a universal constant, $-\lambda$, , at present unknown, without destroying the general covariance"* (Einstein 1917a).[8] The only constraint was that the new constant had to be small enough to make the modification significant only at extremely large distances, in order that the field equations remained compatible with observations of the motion of the planets in our solar system: *"This field equation, with $\lambda$ sufficiently small, is in any case also compatible with the facts of experience derived from the solar system"* (Einstein 1917a).

Einstein then showed that, for the case of the universe as a whole, the modified field equations (9) have the solution

$$\lambda = \frac{\kappa \rho}{2} = \frac{1}{R^2} \qquad (10)$$

where $\lambda$ represents the cosmological constant, $\rho$ is the mean density of matter and $R$ represents the radius of the cosmos (Einstein 1917a). Thus, his model gave a satisfactory relation between the size of the cosmos and the matter it contained. Indeed, in his correspondence around this time, Einstein attempted a rough estimate of the size of the universe (and thus of the cosmological constant) from estimates of the density of matter in the Milky Way, although he later realised that such calculations were unreliable (O'Raifeartaigh *et al.* 2017). He also suggested that a value for the cosmological constant might be estimated by searching for departures from Newtonian predictions in the motion of astrophysical objects (in a manner similar to Seeliger's earlier suggestion), but found the method unsuccessful due to uncertainties in observation (Einstein 1921a, 1921b).[9]

---

[7]Einstein's interpretation of Mach's Principle in these years implied that space could not have a physical reality an infinite distance from matter. This created a puzzle regarding the correct boundary conditions for his cosmology, a puzzle that was only resolved by the postulate of a closed spatial geometry for the universe (Realdi and Peruzzi 2009; Smeenk 2014; O'Raifeartaigh et al. 2017).

[8]It is sometimes stated that Einstein first introduced the cosmological term in a footnote to section 14 of his 1916 'Grundlage' paper (Harvey 2012a; Straumann 2013 p80; Kragh and Overduin 2014 p48). This is not quite correct as Einstein's footnote concerned the Ricci scalar rather than an additional term in the field equations (Einstein 1916 p33).

*3.1  Early interpretations of the cosmological constant*

While the cosmological constant played a key role in Einstein's 1917 model of the cosmos, there is little doubt that an interpretation of the physics underlying the term posed a challenge for him. Indeed, no physical interpretation of the term is presented in the 1917 paper and Einstein's later comments indicate that he viewed $\lambda$ as an uncomfortable mathematical necessity. For example, not long after the publication of his 1917 paper, Einstein remarked to Felix Klein: "*The new version of the theory means, formally, a complication of the foundations and will probably be looked upon by almost all our colleagues as an interesting, though mischievous and superfluous stunt,*[10] *particularly since it is unlikely that empirical support will be obtainable in the foreseeable future. But I see the matter as a necessary addition, without which neither inertia nor geometry are truly relative*" (Einstein 1917c; figure 1). More famously, he declared in 1919: "*But this view of the universe necessitated an extension of equations (1), with the introduction of a new universal constant standing in a fixed relation to the total mass of the universe...... This is gravely detrimental to the formal beauty of the theory*" (Einstein 1919a). Perhaps the best insight into Einstein's view of the term at this time can be found in a rather prescient comment to de Sitter: "*The general theory of relativity allows the addition of the term $\lambda g_{\mu\nu}$ in the field equations. One day, our actual knowledge of the composition of the fixed-star sky, the apparent motions of fixed stars, and the position of spectral lines as a function of distance, will probably have come far enough for us to be able to decide empirically the question of whether or not $\lambda$ vanishes. Conviction is a good mainspring, but a bad judge!*" (Einstein 1917d).

---

[9] Many years later, a similar idea was pursued in order to account for the dynamics of galaxy clusters (Jackson 1970).
[10] The original German reads "*ein … mutwilliges und überflüssiges Kunststück*" (figure 1) which could also be translated as *"a fanciful and superfluous artifice"*.

In March 1918, the Austrian physicist Erwin Schrödinger suggested that a consistent model of a static, matter-filled cosmos could be obtained from Einstein's field equations without the introduction of the cosmological constant (Schrödinger 1918). Essentially, Schrödinger's proposal was that Einstein's cosmic solution could be obtained from the unmodified field equations (5) if a negative-pressure term was added to the energy-momentum tensor on the right-hand side of the equations.[11] Einstein's response was that Schrödinger's formulation was entirely equivalent to that of his 1917 memoir, provided the negative-pressure term was constant (Einstein 1918a).[12] This view may seem surprising, but Einstein's response contained his first physical interpretation of the cosmological term, namely that of a negative mass density: *"In terms of the Newtonian theory…a modification of the theory is required such that "empty space" takes the role of gravitating negative masses which are distributed all over the interstellar space"* (Einstein 1918a). Within a year, Einstein proposed a slightly different interpretation of the cosmic constant, namely that of a constant of integration, rather than a universal constant associated with cosmology: *"But the new formulation has this great advantage, that the quantity appears in the fundamental equations as a constant of integration, and no longer as a universal constant peculiar to the fundamental law"* (Einstein 1919a). Indeed, a letter to Michele Besso suggests that Einstein had arrived at a similar interpretation a year earlier using a variational principle (Einstein 1918b). A follow-up letter to Besso suggests that at one point, Einstein considered the two views to be equivalent: *"Since the world exists as a single specimen, it is essentially the same whether a constant is given the form of one belonging to the natural laws or the form of an 'integration constant'"* (Einstein 1918c).

One explanation for Einstein's ambiguity may be a slight confusion concerning the manner in which the term was introduced. In the opening section of the 1917 paper, Einstein proposed a simple modification of Newtonian gravity: *"In place of Poisson's equation we write $\nabla^2 \phi - \lambda \phi = 4\pi\kappa\rho$ where $\lambda$ denotes a universal constant. If $\rho_0$ be the uniform density of a distribution of mass, then $\phi = -(4\pi\kappa/\lambda)\rho_0$ is a solution of [this] equation"*. This modification is identical to that suggested by Seeliger (see section 2), but was suggested independently by Einstein[13] as a *"foil for what is to follow"* (Einstein 1917a). However, Einstein's modification of the general field equations in the paper was not *"perfectly analogous"* to his modification of Newtonian gravity, contrary to his claim. As pointed out by several analysts,[14] the modified field equations (9) do not reduce in the Newtonian limit to the Seeliger-Poisson equation (4), but to a different relation given by

$$\nabla^2 \phi + c^2 \lambda = 4\pi G \rho \qquad (11)$$

---

[11] See also (Harvey 2009, 2012b).
[12] Schrödinger also suggested that the pressure term might be time variant (Schrödinger 1918; Harvey 2012b), a suggestion that was too speculative for Einstein (Einstein 1918a).

This error may be significant as regards Einstein's interpretation of the cosmic constant. After all, the later view of the cosmological constant term as representing a tendency for empty space to expand would have been deeply problematic for Einstein in 1917, given his view of Mach's Principle at the time (Smeenk 2014; O'Raifeartaigh et al. 2017). It appears that he was shielded from this interpretation by a slight mathematical error, at least in the early years.

Another strange aspect of Einstein's 1917 memoir was his failure to consider the stability of his static model. After all, equation (10) draws a direct relation between a universal constant $\lambda$, the radius of the universe $R$, and the density of matter $\rho$. But the quantity $\rho$ represented a mean value for the latter parameter, arising from the theoretical assumption of a uniform distribution of matter on the largest scales. In the real universe, one would expect a natural variation in this parameter (giving rise to the formation of structure), raising the question of the stability of the model against perturbations in density. It was later shown that the Einstein World is generally unstable against such perturbations: a slight increase in the density of matter would cause the universe to contract, become more dense and contract further, while a slight decrease in density would result in a runaway expansion (Eddington 1930: Eddington 1933 pp 50-54). It is strange that Einstein did not consider this aspect of his model in 1917; some years later, it was a major reason for rejecting the model, as described below.

*3.2 The de Sitter universe*
In July 1917, the Dutch astronomer and theorist Willem de Sitter noted that Einstein's modified field equations allowed an alternate cosmic solution, namely the case of a universe with no matter content (de Sitter 1917). Approximating the known universe as an empty universe, de Sitter set the energy-momentum tensor in Einstein's modified field equations (9) to zero according to

$$G_{\mu\nu} - \frac{1}{2}g_{\mu\nu}G - \lambda g_{\mu\nu} = 0 \qquad (12)$$

and showed that these equations have the solution

$$\rho = 0; \; \lambda = \frac{3}{R^2} \qquad (13)$$

a result he dubbed 'Solution B' to Einstein's 'Solution A' (de Sitter 1917). In this cosmology, Einstein's matter-filled three-dimensional universe of closed spatial geometry was replaced by an empty four-dimensional universe of closed *spacetime* geometry.

---

[13] In a paper of 1919 (Einstein 1919b), Einstein remarked that he was unaware of Seeliger's modification of Newtonian gravity when writing his cosmological memoir in 1917. He cited Seeliger scrupulously after this point (Einstein 1918d p123; Einstein 1931a; Einstein 1933).
[14] See for example (Rindler 1969 p223; Norton 1999; Harvey and Schucking 2000; Earman 2001).

Not surprisingly, Einstein was greatly perturbed by de Sitter's empty universe. Quite apart from the fact that the model was physically unrealistic, the existence of a vacuum solution for the cosmos was in direct conflict with his understanding of Mach's Principle in these years (Smeenk 2014; O'Raifeartaigh et al. 2017). Eventually, Einstein made his criticisms public in a paper of 1918: *"It appears to me that one can raise a grave argument against the admissibility of this solution…..In my opinion, the general theory of relativity is a satisfying system only if it shows that the physical qualities of space are completely determined by matter alone. Therefore no $g_{\mu\nu}$-field must exist (that is no space-time continuum is possible) without matter that generates it"* (Einstein 1918e). In the same paper, Einstein suggested a technical objection to de Sitter's model, namely that it contained a spacetime singularity.

In the years that followed, Einstein debated the relative merits of 'Solution A' and 'Solution B' with de Sitter and other physicists such as Kornel Lanczos, Hermann Weyl, Felix Klein and Gustav Mie. Although he eventually withdrew his remark concerning a singularity, it is clear that Einstein did not accept the de Sitter solution as a realistic model of the universe throughout this debate (Schulmann et al. 1988 pp 351-352). After much confusion, it was eventually realised that the de Sitter solution is not truly a static solution. Indeed, it was shown that the de Sitter metric can be represented by the simple line element

$$ds^2 = e^{a\sqrt{\lambda}t}(-dx^2 - dy^2 - dz^2) + c^2 dt^2 \qquad (14)$$

where *a* and *λ* are arbitrary constants (Lemaître 1925). Presumably, this aspect of the de Sitter universe would have made Einstein even less trusting of the model; it is telling that he did not cite the de Sitter solution in any of his reviews of cosmology around this time (Einstein 1918d pp 116-118, 1921a, 1922a pp 110-111). It is also interesting to speculate that the existence of a mathematically viable vacuum solution to the modified field equations may have marked the beginning of Einstein's distrust of the cosmological constant.

## 4. The cosmological constant in the 1920s

In 1922, the Russian physicist Alexander Friedman suggested that non-static solutions of the Einstein field equations should be considered in relativistic models of the cosmos (Friedman 1922). Starting from the modified field equations (9) and assuming a positive spatial curvature for the cosmos, he derived the two differential equations

$$\frac{3R'^2}{R^2} + \frac{3c^2}{R^2} - \lambda = \kappa c^2 \rho \tag{15}$$

$$\frac{R'^2}{R^2} + \frac{2R''}{R} + \frac{c^2}{R^2} - \lambda = 0 \tag{16}$$

linking the time evolution of the cosmic radius $R$ with the mean density of matter $\rho$ and the cosmological constant $\lambda$. Demonstrating that the Einstein and de Sitter models were special cases of this general class of solutions, Friedman showed that integration of equation (16) gave the simple relation

$$\frac{1}{c^2}\left(\frac{dR}{dt}\right)^2 = \frac{A - R + \frac{\lambda}{3c^2}R^3}{R} \tag{17}$$

and noted that the magnitude of the cosmological constant $\lambda$ determined whether a matter-filled universe expanded monotonically or expanded and then contracted. Setting $\lambda = 0$ in equation (16) above, Friedman even considered the possibility of a cyclic universe.[15]

---

[15] See (Belenkiy 2012, 2013) for a recent review of Friedman's 1922 model.

Few physicists paid attention to Friedman's time-varying cosmology, possibly because the work was quite technical and made no connection to astronomy. Worse, Einstein publicly criticized the paper on the basis that it contained a mathematical error (Einstein 1922b). When it transpired that the error lay in Einstein's criticism, it was retracted a year later (Einstein 1923a). However, an unpublished draft of Einstein's retraction demonstrates that he did not consider Friedman's cosmology to be realistic: *"to this a physical significance can hardly be ascribed"* (Einstein 1923b).[16]

A few years later, the Belgian physicist Georges Lemaître independently derived differential equations for the radius of the cosmos from Einstein's modified field equations. Aware of astronomical observations of the redshift of light from the spiral nebulae by V.M. Slipher (Slipher 1915, 1917) and of emerging evidence of the extra-galactic nature of the spirals (Hubble 1925), Lemaître suggested that the recession of the nebulae was a manifestation of the expansion of space from a pre-existing static cosmos of radius $R_0 = 1/\sqrt{\lambda}$ (Lemaître 1927).[17] This work also received very little attention at first, probably because it was published in a little-read Belgian journal. Lemaître himself brought the work to Einstein's attention in conversation, only to hear the latter dismiss expanding cosmologies as *"abominable"* (Lemaître 1958). Describing the meeting many years later, Lemaître recalled an impression that Einstein's stance stemmed from a lack of knowledge of developments in astronomy: *"Je parlais de vitesses des nébeleuses et j'eus l'impression que Einstein n'était guère au courant des faits astronomiques"* (Lemaître 1958).

*4.1 Zero-point energy and the quantum vacuum*

In 1911, the great German theorist Max Planck proposed that the lowest value of the energy of an oscillator of frequency $\upsilon$ would not be zero, but $\frac{1}{2}h\upsilon$, where $h$ is Planck's constant (Planck 1911). The concept, known as *Nullpunktsenergie* or zero-point energy, was controversial for some years, but received a solid theoretical foundation with the advent of formal quantum mechanics in the mid 1920s.[18] In 1916, the German chemist Walther Nernst suggested that, assuming empty space was filled with electromagnetic radiation, a zero-point energy associated with the vacuum could prevent the heat death of the universe (Nernst 1916). Noting that the energy associated with this phenomenon would in principle be infinitely large, Nernst estimated a value of $\rho_{vac} = 150 \, g/cm^3$ for the energy density of the cosmic vacuum from the Rayleigh-Jeans law of radiation by imposing a cut-off for frequencies above $10^{20}$ Hz (Nernst 1916; Kragh 2012). It should be noted that Nernst's calculations were based on the concept of the ether; in his view, atoms of the chemical elements appeared out of the fluctuations of the ether, and these atoms and their radioactive decay products would later disappear in the zero-point energy of the ethereal sea (Nernst 1921; Kragh 2012).

---

[16] A detailed account of this episode can be found in (Nussbaumer and Bieri 2009 pp 91-92).

[17] It is now known that Lemaître considered many other expansion models in the draft of this paper, but selected the one closest to empirical observation (Luminet 2013).

[18] See (Kragh 2012) for a review.

In the 1920s, Nernst's idea of a zero-point energy of the vacuum was considered by a number of scientists. In particular, Wilhelm Lenz applied the concept to the Einstein Universe, and found that the phenomenon implied a nonsensical value for the size of the cosmos: *"If one allows waves of the shortest observed wavelengths of λ ≅ 2 × 10⁻¹¹ cm ....and if this radiation, converted to material density (u/c² ≅ 10⁶) contributed to the curvature of the world—one would obtain a vacuum energy density of such a value that the world would not reach even to the moon"* (Lenz 1926; transl. Kragh 2012). A similar calculation was carried out by Wolfgang Pauli, with similar results.[19] Several other physicists considered the issue in these years; in general, the conclusion was that observation implied that the zero-point energy of the vacuum was not a real effect (Jordan and Pauli 1928; Pauli 1933, 1945). As Pauli remarked: *"It is more consistent ...to exclude a zero-point energy for each degree of freedom as this energy, evidently from experience, does not interact with the gravitational field"* (Pauli 1933 p250).[20]

## 5. The cosmological constant in the 1930s: the expanding universe

In 1929, the American astronomer Edwin Hubble published the first evidence of a linear relation between the redshifts of the spiral nebulae and their radial distance (Hubble 1929). The discovery marked a turning point in modern cosmology as the data could not be explained in the context of Einstein's static matter-filled world (solution A) or de Sitter's empty universe (solution B). As de Sitter remarked at a meeting of the Royal Astronomical Society (RAS) in January 1930: *"It would be desirable to know what happens when we insert matter into the empty world represented by solution B. The difficulty in the investigation of*

---

[19] Pauli's calculation was never published but is described in (Enz and Thellung 1960) and has been reconstructed in (Straumann 1999, 2002).
[20] See (Rugh and Zinkernagel 2002) or (Kragh 2012) for further details.

*this problem lies in the fact that it is not static"* (de Sitter 1930a). A report of the meeting was read by Georges Lemaître and, following a communication from him, Arthur Stanley Eddington arranged for Lemaître's 1927 paper to be republished in English translation in the Monthly Notices of the RAS, bringing the work to a wider audience (Lemaître 1931a).[21] Soon, a number of papers had emerged that explored expanding models of the Friedman-Lemaître type for diverse values of cosmic parameters such as spatial curvature and the cosmological constant (Eddington 1930, 1931a: de Sitter 1930b, 1930c; Einstein 1931a; Einstein and de Sitter 1932; Tolman 1930, 1931a, 1932; Heckmann 1931, 1932; Robertson 1932).[22]

*5.1 Einstein abandons the cosmic constant*

Einstein was one of the first to accept Hubble's observations as likely evidence of a non-static universe, as evidenced by several statements he made during a sojourn in California in early 1931. For example, the *New York Times* reported Einstein as commenting that *"New observations by Hubble and Humason concerning the redshift of light in distant nebulae make the presumptions near that the general structure of the universe is not static"* (AP 1931a) and *"The redshift of the distant nebulae have smashed my old construction like a hammer blow"* (AP 1931b). In April 1931, Einstein published a model of the expanding cosmos based on Friedman's 1922 analysis of a matter-filled dynamic universe of positive spatial curvature (Einstein 1931a).[23] The most important feature of this model, sometimes known as the Friedman-Einstein model, was that Einstein dispensed with the cosmological constant term, for two stated reasons. First, the term was unsatisfactory because it did not provide a stable static solution: *"It can also be shown… that this solution is not stable. On these grounds alone, I am no longer inclined to ascribe a physical meaning to my former solution"* (Einstein 1931a). Second, the term was unnecessary because the assumption of stasis was not justified by observation: *"Now that it has become clear from Hubbel's* [sic] *results that the extra-galactic nebulae are uniformly distributed throughout space and are in dilatory motion (at least if their systematic redshifts are to be interpreted as Doppler effects), assumption (2) concerning the static nature of space has no longer any justification"* (Einstein 1931a). Indeed, an early portend of this strategy can be found in a note written by

---

[21] See (Nussbaumer and Bierri 2009 pp 121-122) for a description of this episode.

[22] See (de Sitter 1932 pp 121-128) or (Robertson 1933) for a contemporaneous review.

[23] We have recently published an analysis and English translation of this work (O'Raifeartaigh and McCann 2014).

Einstein to Hermann Weyl in 1923. In the course of a discussion of the de Sitter model, Einstein wrote: *"If there is no quasi-static world after all, then away with the cosmological term"* (Einstein 1923c; Straumann 2002; Nussbaumer and Bieri 2009 pp 82-83).

Setting the cosmological term to zero in Friedman's analysis of 1922, Einstein derived simple expressions relating the rate of cosmic expansion to key parameters such as the present radius of the cosmos, the mean density of matter and the timespan of the expansion. Using Hubble's empirical estimate of 500 km s$^{-1}$Mpc$^{-1}$ for the recession rate of the nebulae, he then calculated numerical values of $10^8$ light-years, $10^{-26}$ g/cm$^3$ and $10^{10}$ years for each of these parameters respectively (Einstein 1931a). We have previously noted that these calculations contain a slight systematic numerical error; the Hubble constant above in fact implied a value of 2x10$^9$ light-years, $10^{-28}$ g/cm$^3$ and 2x10$^9$ years for the radius of the cosmos, the mean density of matter and the timespan of the expansion respectively (O'Raifeartaigh and McCann 2014). However the error was a small one and did not substantially affect a major puzzle raised by the model; if the timespan of cosmic expansion represented the age of the universe, it was strangely small in comparison with estimates of the age of stars (as calculated from astrophysics) or estimates of the age of the earth (as deduced from radioactivity). Einstein attributed this age paradox to the idealized assumptions of the model, in particular the assumption of a homogeneous distribution of matter on the largest scales (Einstein 1931a).

In 1932, Einstein collaborated with Willem de Sitter to propose an even simpler model of the expanding universe. Following an observation by Otto Heckmann (Heckmann 1931, 1932) that the presence of matter in a non-static universe did not necessarily imply a positive curvature of space, and mindful of a lack of empirical evidence for spatial curvature, Einstein and de Sitter set both the cosmological constant and spatial curvature to zero (Einstein and de Sitter 1932). An intriguing facet of this model was that the rate of expansion and the density of matter $\rho$ were related by the simple equation

$$\left(\frac{R'}{R}\right)^2 = \frac{1}{3}\kappa\rho c^2 \qquad (18)$$

where $\kappa$ is the Einstein constant. Applying Hubble's empirical value of $H_0 = 500$ km s$^{-1}$ Mpc$^{-1}$ for the recession rate of the galaxies, the authors found that it predicted a value of 4x10$^{-28}$ g cm$^{-3}$ for the mean density of matter in the cosmos, a prediction they found reasonably compatible with contemporaneous estimates from astronomy. However, it is interesting to note that the authors were careful in their paper not to dismiss the possibility of spatial curvature: *"We must conclude that at the present time it is possible to represent the facts without assuming a curvature of three-dimensional space. The curvature is, however, essentially determinable, and an increase in the precision of the data derived from observations will enable us in the future to fix its sign and to determine its value"* (Einstein and de Sitter 1932). No such courtesy was afforded to the cosmological constant.

The Einstein-de Sitter model became very well-known and went on to play a significant role in 20$^{th}$ century cosmology. One reason was that it marked an important hypothetical case in which the expansion of the universe was precisely balanced by a critical density of matter, given by equation (18) as $\rho_c = 3H_0^2/8\pi G$. This allowed for a useful classification of cosmic models; assuming a vanishing cosmological constant, a cosmos of mass density higher than the critical value would be of spherical geometry and eventually collapse, while a cosmos of mass density less than the critical value would be of hyperbolic spatial geometry and expand at an ever increasing rate. Indeed, the geometry of such models is usefully described in terms of the 'density parameter' Ω, defined as the ratio of the actual matter density of the universe $\rho$ to the critical density $\rho_c$ required for spatial closure, i.e. $\Omega = \rho/\rho_c$. This simple classification scheme could be generalized to models with a cosmological constant by defining the energy density parameter as $\Omega = (\Omega_M + \Omega_\lambda)$, where $\Omega_M$ and $\Omega_\lambda$ represented the energy density contributions due to matter and the cosmological constant respectively. In this scheme, the Einstein-de Sitter universe is neatly specified as ($\Omega = 1: \Omega_M = 1, \Omega_\lambda = 0$), while the empty de Sitter universe is described as ($\Omega = 1: \Omega_M = 0, \Omega_\lambda = 1$). We note that it is easily shown that $\Omega_M = (8\pi G/3H_0^2)\rho_M$ and $\Omega_\lambda = \lambda/3H_0^2$.

The Einstein-de Sitter model also marked an important benchmark case for observers; in the absence of empirical evidence for spatial curvature or a cosmological constant, it seemed the cosmos could be described in terms of just two parameters, each of which could be determined independently by astronomy. Indeed, the theory became the standard cosmic model for astronomers for many years, although it suffered from a similar timespan problem as the Friedman-Einstein model.[24] In addition, empirical observations soon suggested a mean cosmic density of matter far below the critical value, as will be discussed below.

---

[24] Although it was not considered in the original paper, it is easily shown that the timespan of expansion of the Einstein-de Sitter model is given as *t* = 2/(3*H*$_0$) (Einstein 1933).

It is clear from the above that Einstein lost little time in abandoning the cosmological constant when presented with empirical evidence for a non-static universe.[25] Certainly, he was never to re-instate the term in the field equations after 1931 and he is even reputed to have described the term in later years as *"my biggest blunder"*. Whether Einstein used these exact words may never be known[26] but if he did, it is likely that he was referring to his failure to consider the stability of his static cosmology of 1917. Perhaps the best indication of Einstein's ultimate view of the cosmic constant can be found in his 1945 review of relativistic cosmology: *"If Hubble's expansion had been discovered at the time of the creation of the general theory of relativity, the cosmologic member would never have been introduced. It seems now so much less justified to introduce such a member into the field equations, since its introduction loses its sole original justification"* (Einstein 1945 p130).

*5.2 Others retain the cosmological constant*

At first, not many of Einstein's colleagues took his lead in abandoning the cosmological constant. Some felt that the term should be retained for reasons of mathematical generality; others felt that it could be used to address cosmological puzzles such as the timespan of the expansion and the formation of galaxies in an expanding universe. Still others felt that the term had an important role to play in giving a physical cause for cosmic expansion.

Considering the mathematical argument first, many theoreticians noted that the modified equations (9) represent the most general form of the field equations. For example, to satisfy the conservation of energy-momentum, the tensor representing the space-time metric must have a vanishing divergence. The most general second-order tensor that satisfies this criterion is not $G_{\mu\nu} - \frac{1}{2}g_{\mu\nu}G$, the left-hand side of equation (6), but $G_{\mu\nu} - \frac{1}{2}g_{\mu\nu}G + \lambda g_{\mu\nu}$, the left-hand side of equation (9). As noted in section 3, empirical observation demanded that $\lambda$ was extremely small but from a theoretical point of view, there was no reason that it should be exactly zero. Thus, some theoreticians felt it was an error to assign the value zero to a term whose value was in fact unknown. This viewpoint was neatly expressed by Richard Tolman in a letter to Einstein in September 1931: *"…since the introduction of the Λ–term provides the most general possible expression of the second order which would have the right*

---

[25] With one exception, as will be discussed in section 6.

[26] The statement was reported by the Russian physicist George Gamow (Gamow 1956; Gamow 1970 p44). Some doubt has been cast on the accuracy of Gamow's report in recent years (Straumann 2002; Livio 2013 pp 231-243), while the report has been supported by Ralph Alpher (Topper 2013 p165) and by John Archibald Wheeler (Taylor and Wheeler 2000 p G-11).

*properties for the energy-momentum tensor, a definite assignment of Λ=0, in the absence of experimental determination of its magnitude, seems arbitrary and not necessarily correct"* (Tolman 1931b).[27] A similar view can be found in Willem de Sitter's 1932 review of relativistic cosmology: *"As a matter of fact, neither the average density nor the rate of expansion are at the present time known with sufficient accuracy to make an actual determination [of λ] possible....All we can say is that, if the curvature is small, then λ must be small, and if the curvature is very small, then λ must be very small"* (de Sitter 1932 p127).

A second argument for the retention of the cosmic constant arose from considerations of the timespan of the cosmic expansion. As noted above, with the cosmological constant set to zero in Friedman's analysis, Hubble's observations implied a time of expansion of about 2 billion years, a figure that was strangely small in comparison with contemporaneous estimates of the age of stars and the age of the earth. Several physicists suggested that the cosmological constant could play a role in resolving the paradox. For example, ~~Arthur Stanley~~ Eddington noted that a positive cosmic constant could give a model in which the cosmos expanded from a static universe of indefinite age (Eddington 1930, 1931a). This model, a more detailed version of Lemaître's analysis of 1927 (Lemaître 1927, 1931a), became known as the Eddington-Lemaître model. By this time, Lemaître himself had proposed his famous hypothesis of a universe that originated as a 'primeval atom' (Lemaître 1931b; Kragh and Lambert 2007). With such cosmic origins in mind, he noted that a judicious choice of value for the cosmological constant could give a cosmic expansion in three stages; an initial phase during which the expansion is de-accelerated by gravity, a 'loitering' phase in which the de-acceleration is balanced by the repulsive influence of the cosmic constant, and a final phase in which the repulsion becomes dominant (Lemaître 1931c, 1931d, 1933). Here the cosmic expansion was governed by a cosmological constant given by $\lambda = \lambda_E (1 + \epsilon)$, where $\lambda_E$ was the value of the cosmic constant in Einstein's static model of 1917 and the adjustable parameter $\epsilon$ determined the length of the stagnation period. A schematic of this model, known as the 'hesitating' or 'loitering' universe, is shown in figure 2 along with the Eddington-Lemaître model.

A promising facet of Lemaître's hesitating model was that it also offered a possible mechanism for the formation of galactic structures, a phenomenon that presented a formidable puzzle in the context of the discovery of cosmic expansion (Kragh 1996 pp 288-

---

[27] As far as we know, Tolman was the first to use the symbol Λ for the cosmological constant. We shall use the symbols λ and Λ interchangeably as they were used by various authors.

289). In Lemaître's view, the temporary balance of the gravitational force and the cosmological constant during the stagnant phase of the hesitating universe presented a stable interval during which perturbations in matter density could condense into galaxies and galaxy clusters (Lemaître 1931c, 1931d, 1933, 1934). Indeed, such a process could then give rise to a third phase in which the cosmic constant dominated. This model attracted some attention, especially when combined with later work on the formation of nebulae in the early universe (Gamow and Teller 1939a, 1939b). For example, in his 1952 exposition of cosmology, the well-known theorist Hermann Bondi commented: *"Lemaître's model….has many attractive features and, especially if combined with the work of Gamow and Teller, seems to be the best relativistic cosmology can offer. The timescale difficulty is largely resolved through the interposition of the arbitrarily long 'quasi-Einstein' stage"* (Bondi 1952 p121).

A fourth argument for the retention of the cosmic constant arose from considerations of the physical *cause* of cosmic expansion. As pointed out by Eddington, relativity allowed for an expanding universe, but it did not *explain* the phenomenon. In Eddington's view, the cosmological constant supplied a physical explanation for the phenomenon: *"It is found similarly that the added term ($\lambda g_{\mu\nu}$) gives rise to a repulsion directly proportional to the distance…It is a dispersive force like that which I imagined as scattering apart the audience in the lecture-room* (Eddington 1933 p23). A similar point was made by Willem de Sitter, who asked: *"What is it then that causes the expansion? Who blows up the india-rubber ball? The only possible answer is: the lambda does it"* (de Sitter 1931). Thus, by the early 1930s, the cosmic constant was seen as a repulsive force arising from a negative pressure (Maneff 1932; Zaycoff 1932; Robertson 1933). More specifically, Georges Lemaître associated the term with a pressure arising from an energy density of the vacuum and gave a rough estimate of the magnitude of the effect from observational constraints: *"Everything happens as though the energy in vacuo would be different from zero. In order that absolute motion, i.e., motion relative to vacuum, may not be detected, we must associate a pressure $p = -\rho c^2$ to the density of energy $\rho c^2$ of vacuum. This is essentially the meaning of the cosmical constant $\lambda$ which corresponds to a negative density of vacuum $\rho_0$ according to $\rho_0 = \lambda c^2/4\pi G \cong 10^{-27} g/cm^3$"* (Lemaître 1934).[28] A similar idea was carried further by the Soviet physicist Matvei Bronstein, who proposed a model involving a continuous transfer of energy between ordinary matter and the energy of the vacuum (Bronstein 1933). This proposal envisioned a time-varying cosmological constant, a prediction that can be seen as a precursor of the modern concept of quintessence.[29]

---

[28] We note that Lemaître associated a positive cosmological constant with a *negative* energy density of space and did not draw a connection with the zero-point energy of the vacuum at this point (Kragh and Overduin 2014 p52).

It should be noted that Eddington had an additional motivation for the retention of the cosmological constant, namely the use of the term in attempts to construct a more general theory that unified general relativity with quantum theory (Eddington 1931b: Kragh 2015). Although Eddington's 'fundamental theory' is now of historical interest only, the theory serves as an example of yet another role for the cosmological constant; from the very beginning, the term found application in attempts to connect the general theory of relativity with grander theories (Weyl 1918; Einstein 1919a), an application that pertains to this day.[30]

Thus, in the years following the discovery of the expanding universe, some theoreticians felt that the cosmic constant term had an important role to play in relativistic cosmology. As Eddington remarked in his classic book '*The Expanding Universe*': *"I would as soon think of reverting to Newtonian theory as of dropping the cosmological constant"* (Eddington 1933, p24). Perhaps the best summary of this approach was given by Georges Lemaître some years later in his contribution to an Einstein Festschrift of 1949: *"Even if the introduction of the cosmological constant has lost its original justification… it remains true that Einstein has shown that the structure of his equations quite naturally allows for the presence of a second constant besides the gravitational one….the history of science provides many instances of discoveries which have been made for reasons which are no longer satisfactory. It may be that the discovery of the cosmological constant is such a case"* (Lemaître 1949 p443).[31]

*5.3 Advances in analysis*

The discovery of the recession of the nebulae also spurred the development of formal tools for the description of cosmological models. In particular, it was shown (Robertson 1935; Walker 1937) that all time-dependent cosmologies that assume a spacetime that is isotropic and homogeneous (see section 3) can be described by the generic space-time metric

$$ds^2 = c^2 dt^2 - R^2(t) d\sigma^2 \qquad (19)$$

---

[29] See (Kragh 2012) for further details of Bronstein's model.

[30] See (McCrea 1971; Ray 1990) for further discussion of this point.

[31] In this article, Lemaître connects λ with zero-point energy, stating "*..it is necessary the theory should provide some possibility of adjustment when the zero-level from which energy is counted, is changed arbitrarily*". See (Seitter and Duemmler 1989) for further discussion.

In this expression, $R(t)$ represents a time-dependent scale factor of expansion and $d\sigma^2$ is a line element of Riemannian space given by

$$d\sigma^2 = \left(\frac{dr^2}{1-kr^2} + r^2(d\theta^2 + sin^2\theta d\phi^2)\right) \quad (20)$$

where $r$, $\theta$ and $\phi$ are co-moving co-ordinates and $k$ is a curvature parameter normalized to assume the values +1, 0 or -1 for positive, zero or negative spatial curvature respectively.[32] In terms of astronomical observation, the rate of cosmic expansion is related to the cosmological redshift according to the simple relation

$$cz = H_0 D_0 \quad (21)$$

where $z$ represents the fractional change in wavelength $\Delta\lambda/\lambda_0$ of the light from a nebula, the Hubble constant $H_0$ is the fractional rate of change of the scale factor $R'/R$ and $D_0$ is an appropriately defined distance measure. In astronomical practice, the redshifts of the nebulae were found to conform well to a relation of the form

$$m = 5log_{10}z + X \quad (22)$$

where $m$ represented the apparent magnitude of a nebula (a measure of its distance) and X was a constant (Hubble and Humason 1931; Bondi 1952 p39). We note that, as telescopes reached out to larger and larger distances, the question arose as to whether the Hubble constant evolved over time. In particular, it was expected that the rate of cosmic expansion would be slowed by the self-gravity of galaxies, as will be discussed in section 7.

## 6. The cosmological constant in the 1940s

In the 1940s, few physicists outside the relativity community paid attention to the Lemaître- or Eddington-Lemaître models of the universe. While many accepted Hubble's observations as possible evidence for an expanding universe, theories concerning cosmic origins were considered deeply speculative, an attitude that persisted for some years (Kragh 1996 pp 135-143). Interest focused instead on the possibility of determining more accurate estimates of cosmic parameters such as the Hubble constant, the mean density of matter and the curvature of space using large telescopes, but the construction of such instruments was delayed by the second world war (Longair 2006 pp 118-120). On the other hand, two theoretical advances occurred in this decade that were to have a major bearing on the story of the cosmological constant.

---

[32] See (Kolb and Turner 1990 pp 29-30) or (Weinberg 1972 pp 412-413) for further discussion of the Robertson-Walker metric.

*6.1 The hypothesis of primordial nucleosynthesis*

In the late 1940s, the Russian émigré physicist George Gamow suggested that Friedman-Lemaître cosmologies might offer a radical solution to the puzzle of nucleosynthesis. With the failure of standard models of stellar nucleosynthesis to explain the relative abundance of the lightest chemical elements, Gamow and his colleagues Ralph Alpher and Robert Herman explored whether the phenomenon might be described in the context of nuclear processes in a young universe that was once extremely dense and hot (Gamow 1942, 1946; Alpher, Bethe and Gamow 1948; Alpher and Herman 1948, 1950, 1951). While this hypothesis did not have any direct bearing on the cosmological constant at first,[33] it opened up a new line of enquiry for dynamic cosmologies ~~and~~ that was later to set important constraints on estimates of the mean density of matter in the cosmos, as will be discussed below.

*6.2 The cosmological constant in steady-state cosmology*

In parallel with the work of Gamow et al., a new type of cosmic model was proposed in the United Kingdom known as the 'steady-state' universe. In this cosmology, the universe expands but remains essentially unchanged in every other respect. Today, the steady-state universe is mainly associated with the Cambridge physicists Fred Hoyle, Hermann Bondi and Thomas Gold, but other theorists entertained similar ideas.

In particular, it has recently been discovered that, soon after the publication of Hubble's graph of 1929, Einstein himself briefly considered the notion of an expanding universe in a steady state. In an unpublished work, Einstein proposed that the density of matter could be maintained constant in an expanding universe by a continuous formation of matter from empty space, a process he attributed to an energy associated with the cosmological constant: *"If one considers a physically bounded volume, particles of matter will be continually leaving it. For the density to remain constant, new particles of matter must be continually formed within that volume from space. The conservation law is preserved in that, by setting the λ-term, space itself is not empty of energy; its validity is well known to be guaranteed by equations (1)"* (Einstein 1931b). Indeed, Einstein proposed the process as a possible cause of cosmic expansion: *"The density is therefore constant and determines the expansion apart from its sign"* (Einstein 1931b). However, Einstein soon found that his steady-state model contained a fatal flaw and he abandoned the idea before publication.[34]

---

[33] The group generally ignored the cosmological constant in their analysis, although Gamow employed the term to address the puzzle of the timespan of expansion in a wide-ranging paper of 1949 (Gamow 1949).

In the late 1940s, Hoyle, Bondi and Gold became sceptical of Lemaître's idea of a fireworks origin for the universe and noted that evolving cosmologies predicted an age for the universe that was problematic. In consequence, the trio explored the idea of an expanding universe that remains essentially unchanged due to a continuous creation of matter from the vacuum. For Bondi and Gold, the idea followed from their belief in the 'perfect cosmological principle', a philosophical principle that proposed that the universe should appear essentially the same to observers in all places *at all times* (Bondi and Gold 1948). We shall not discuss their model further as it was not formulated in the context of the general theory of relativity. By contrast, Fred Hoyle constructed a steady-state model of the cosmos by modifying the field equations (6) according to

$$\left(G_{\mu\nu} - \frac{1}{2}g_{\mu\nu}G\right) + C_{\mu\nu} = -\kappa\, T_{\mu\nu} \qquad (23)$$

where the tensor $C_{\mu\nu}$ was a 'creation-field' term representing the continuous creation of matter from the vacuum (Hoyle 1948). In many ways, the new term acted like a positive cosmological constant, giving an exponential expansion of space. Indeed, the line element of the Hoyle model can be written as

$$ds^2 = e^{a\sqrt{\lambda}t}(-dx^2 - dy^2 - dz^2) + c^2 dt^2 \qquad (24)$$

---

[34] See (O'Raifeartaigh et al. 2014; Nussbaumer 2014b) for further details on Einstein's attempt at a steady-state model.

almost identical to that of the de Sitter model.[35] As Hoyle remarked: *"The $C_{\mu\nu}$ term ..plays a role similar to that of the cosmical constant in the de Sitter model, with the important difference that, however, there is no contribution from the $C_{00}$ component…..this difference enables a universe, formally similar to the de Sitter model to be obtained, but in which ρ is non-zero"* (Hoyle 1948). A few years later, the British physicist William McCrea proposed a slightly different formulation of Hoyle's model, in which the 'creation-field' was replaced by a scalar field on the right-hand side of the field equations, representing a negative pressure (McCrea 1951; Kragh 1996 pp 205-206). This term could be represented by the equation of state $p = -\rho c^2$, exactly as in the case of Lemaître's model of 1934 (see section 5.2).

As is well known, a significant debate developed during the 1950s and 1960s between proponents of the steady-state and evolving models of the cosmos. Eventually, steady-state models were effectively ruled out by astronomical discoveries such as the distribution of galaxies at different epochs and the detection of the cosmic microwave background (Kragh 1996 pp 318-338). However, several aspects of steady-state models – in particular the use of the de Sitter metric and its association with the cosmological constant – have found new relevance in the context of the theory of cosmic inflation, as will be discussed in section 8.

## 7. The cosmological constant in the years 1950-1970

The hypothesis of steady-state cosmology spurred new efforts to determine key cosmological parameters by astronomical observation. In particular, the opening of the 200-inch Hale telescope at the Palomar Observatory in California in 1949 heralded a new era of practical cosmology. In this work, attention focused on the Einstein-de Sitter model as it could be characterized by just two parameters, the current rate of cosmic expansion $H_0$ and the current mean density of matter $\rho_0$. Indeed, the challenge to establish observational values for these parameters was later dubbed "the search for two numbers" (Sandage 1961, 1970).

In the first instance, ground-breaking observations by the American astronomers Walter Baade and Allan Sandage led to a successive recalibration of the distance to the galaxies (Baade 1952; Sandage 1958); this recalibration suggested a smaller Hubble constant, implying a longer timespan of cosmic expansion. By the end of the 1950s, Hubble's original estimate of $H_0$= 500 km s$^{-1}$ Mpc$^{-1}$ had been reduced to 75 ± 25 km s$^{-1}$ Mpc$^{-1}$, implying a timespan of the order of 8 billion years even for models without a cosmological constant (Sandage 1961). However, estimates of the age of the oldest stars in our galaxy were now of the order of 15 billion years, despite large uncertainties, suggestive of a new conundrum concerning the time of expansion. It is interesting to note that Allan Sandage, the acknowledged leader of this epoch of observational cosmology, suggested in 1961 that this conflict might be indicative of a need for a positive cosmological constant in evolving models of the cosmos (Sandage 1961).

---

[35] It is easily shown that the assumption of a continuous creation of matter necessitates this metric (Hoyle 1948; Weinberg 1972 pp 459-460).

Meanwhile, estimates of the mean density of matter in the universe suggested a value far below the critical value of the Einstein-de Sitter model. One method was to count the number of galaxies in a given volume of space and multiply by the mass of each galaxy, the latter figure being obtained by measuring the average luminosity of galaxies and converting to mass by means of the mass-to-light ratio of galactic matter (Longair 2006 pp 357-360). A second method was to determine the mass of galaxies by dynamical methods, i.e., by comparing the motion of rotating galaxies with that predicted by Kepler's laws. While both methods entailed large uncertainties, the dynamical method led to the discovery that, on the scale of galaxies and galaxy clusters, a large portion of matter takes the form of dark matter, i.e., is detectable only by its gravitational effect.[36]

An important new approach to measuring the density of matter was suggested in the mid-1950s, namely to search for an expected slowing in the rate of cosmic expansion over time (Robertson 1955; Humason, Mayall and Sandage 1956; Hoyle and Sandage 1956). From the Friedman equations (15) and (16), it is easily shown that for models without a cosmological constant, one might expect a slowing in the rate of cosmic expansion due to the presence of matter according to

$$\frac{R_0''}{R_0} = -\frac{4\pi G}{3}\rho_0 \qquad (25)$$

where the subscripts represent present values for cosmic parameters. Defining a de-acceleration parameter $q_0$ as

$$q_0 = -\frac{1}{H_0^2}\frac{R_0''}{R_0} \qquad (26)$$

---

[36] See (Ostriker and Mitton 2013 pp 174-197) for a review.

it follows that

$$q_0 = \frac{4\pi G}{3H_0^2}\rho_0 \tag{27}$$

In terms of the density parameter $\Omega_M$ (section 5), it follows that

$$q_0 = \frac{1}{2}\Omega_M \tag{28}$$

or, more generally,

$$q_0 = \frac{1}{2}\Omega_M - \Omega_\Lambda \tag{29}$$

for cosmologies with a non-zero cosmological constant.

Thus a measurement of the de-acceleration parameter $q_0$ could yield a direct estimate of the density of matter, at least for models without a cosmological constant. We also note that one could expect a value of $q_0 > ½$ for a cosmos of closed spatial curvature, a value of $q_0 < ½$ for a cosmos of open geometry and a value of $q_0 = ½$ for a universe of Euclidean geometry. Most importantly, a value of $q_0 = -1$ was predicted for steady-state models,[37] raising the prospect of a clear distinction between steady-state and evolving cosmologies based on measurements of de-acceleration (Hoyle and Sandage 1956).

From the point of view of observation, the de-acceleration parameter $q_0$ could in principle be determined by measuring redshift/distance relations for galaxies at great distance and comparing the results with that for more local galaxies. In this work, the simple magnitude-redshift relation (22) was replaced by equations of the type

$$m = 5\log cz + 1.086(1 - q_0)z + \cdots \tag{30}$$

where the parameters *m* and *z* represent the apparent magnitude and fractional redshift of a nebula respectively (Robertson 1955; Hoyle and Sandage 1956). Thus, an ambitious astronomical programme was undertaken to determine [log *z*, *m*] plots for galaxies at great distance, a modified version of the "search for two numbers" (Humason, Mayall and Sandage 1956; Sandage 1961, 1970; Longair 2006 pp 349-350). The results of this program suggested a de-acceleration in cosmic expansion of the order of $q_0 \sim 1.2 \pm 0.4$, an estimate that seemed to favour the Friedman models over the steady-state solution (Sandage 1970). However, this result was subject to large uncertainties, not least because the program was limited to galaxies with redshifts smaller than $z = 0.5$ and because little account was taken of galactic evolution. One intriguing aspect of the program was a prescient proposal that the study be extended to larger redshifts with the use of type Ia supernovae as standard candles (Tammann 1979; Tammann, Sandage and Yahil 1979).

---

[37] The Hubble constant does not vary over time in steady-state models (Hoyle 1948).

*7.1 The redshifts of the quasars*

The 1960s also saw the discovery of puzzling new astronomical entities, extremely luminous objects apparently lying at tremendous distance (Schmidt 1963, 1965: Schmidt and Matthews 1964; Sandage 1965). Two unusual aspects of the entities (soon known as quasi-stellar objects or 'quasars') were that they did not appear to conform to the standard relation between redshift and distance obeyed by ordinary galaxies, while they exhibited a preponderance of redshifts at around the large value of $z = 2$ (Hoyle and Burbidge 1966; Longair and Scheuer 1967; Burbidge and Burbidge 1967). The discovery prompted a new appraisal of Lemaître's hesitating model of the cosmos (see section 5.2), with a number of physicists interpreting the phenomenon as evidence for a stagnant phase in cosmic expansion due to a positive cosmological constant (Petrosian, Salpeter and Szekeres 1967; Shklovsky 1967; Kardashev 1967; Rowan-Robinson 1968; Petrosian and Salpeter 1970). Indeed, in a detailed analysis of quasar observations in the context of Lemaître's model, the Russian astrophysicist Nikolai Kardashev calculated a value of $2 \times 10^{-5}$ for Lemaître's stagnation parameter $\epsilon$ (Kardashev 1967), while this calculation was corrected on further analysis to $\epsilon = 6 \times 10^{-5}$ by the British astronomer Michael Rowan-Robinson (Rowan-Robinson 1968).

In time, the association of the redshifts of the quasars with a stagnant phase in cosmic expansion fell from favour, as quasars with ever larger redshifts were subsequently detected (Petrosian 1974). Indeed, the latter discovery effectively ruled out the hesitating model, side-lining the cosmological constant once again. As the American astronomer Vahe Petrosian remarked in a substantial review of this episode at the 1973 meeting of the International Astronomical Union: "*In the absence of strong evidence in favour of Lemaître models, we must again send back the Lemaître models and along with them the cosmological constant until their next reappearance*" (Petrosian 1974).

One important outcome of this episode was that it prompted the eminent Russian theorist Yakov Zel'dovich (figure 3) to re-examine the question of a non-zero cosmological constant, and the physics underlying such a term: *"To what extent was the assumption of Λ=0, which was frequently made recently justified?.... The genie has been let out of the bottle and it is no longer easy to force it back in. Even if Λ = 0 exactly, it is now necessary to arrive at this answer with great difficulty, slowly, gradually, by narrowing the ranges…. We witness the birth of a new field of activity, namely the determination of Λ. But first let us answer the following question; how is it possible to visualise the meaning of the cosmological constant? Why is its determination interesting for physics as a whole?"* (Zel'dovich 1968). These questions marked an important point in the history of the cosmological constant, as it was seen for the first time as an issue of wide significance in physics.

Zel'dovich's starting point was a re-analysis of Lemaître's proposal of an energy contribution from the vacuum, this time in the context of modern quantum field theory. First, he noted that the scalar field associated with the quantum zero-point energy of the vacuum takes the form of an effective cosmological constant (Zel'dovich 1967, 1968). He then found that basic principles of quantum field theory suggested a lower bound of $10^{-10}$ cm$^{-2}$ for the cosmological constant (corresponding to an energy density of $\rho_\Lambda = 10^{17}$ g/cm$^3$), far exceeding a maximum value of $\Lambda = 10^{-54}$ cm$^{-2}$ (or $\rho_\Lambda = 10^{-29}$ g/cm$^3$) set by observation. This calculation constituted a more quantitative formulation of the puzzle noted by Lenz, Jordan and Pauli (section 4.1); in addition, the demonstration of the Casimir effect in the late 1950s (Casimir 1948; Sparnaay 1957) had convinced many physicists of the reality of the zero-point energy of the vacuum.[38] Thus Zel'dovich's calculations were taken seriously; that quantum field theory predicted an effect that was forty orders of magnitude larger than that observed was a theoretical puzzle that became known as the '*cosmological constant problem*'. One possible solution was a modification of the general field equations by means of an additional scalar field that could cancel the zero-point energy, although such a process seemed implausible as the latter involved fluctuations of many fields associated with a myriad of particles and their interactions (Zel'dovich 1968). However, the problem was not yet considered too pressing; given the lack of empirical evidence for a cosmological constant, most physicists assumed that the quantum energy of the vacuum was reduced to zero by some as-yet unknown symmetry principle (Weinberg 1989).

---

[38] In particular, quantum field theory predicts that the energy contribution of virtual particles and fields must be taken into account. See (Abbott 1988) for a non-technical overview.

# 8. The cosmological constant in the years 1970-1990

By the late 1960s, the steady-state hypothesis had been effectively marginalized by observations of the distribution of the galaxies at different epochs and the detection of the cosmic background radiation.[39] Meanwhile, astronomical observations appeared to be reasonably consistent with the simplest evolving models of the cosmos, with some anomalies. However, these anomalies became more significant as the years progressed.

*8.1 The problem of missing mass*

As astronomers probed further and further into the distant universe in the 1970s, new estimates of the mass density of the universe using the methods of galaxy counting and of rotational dynamics continued to suggest an upper limit far below the critical value of the Einstein-de Sitter model. This data appeared to conflict with a continuing lack of evidence of spatial curvature and with contemporaneous estimates of the Hubble constant (see (Gott et al. 1974) for a contemporaneous review). In addition, a universe of flat geometry was favoured on theoretical grounds, a puzzle that became known amongst theorists as the *flatness problem*.[40]

*8.2 Galactic evolution and $q_0$*

In the mid-1970s, the British-born astronomer Beatrice Tinsley embarked on a detailed study of galactic evolution. A startling outcome of this programme was the suggestion that previous estimates of the de-acceleration parameter $q_0$ (see section 7) had not taken sufficient account of the evolution of galaxies, resulting in a significant overestimate of this parameter. By the mid-70s, spectrophotometric observations of very distant galaxies hinted at a value for $q_0$ significantly smaller than previous estimates (Gunn and Oke 1974; Gunn 1975). Indeed, when corrected for galactic evolution, these data were interpreted by some as evidence for a *negative* de-acceleration parameter, i.e., for a positive cosmological constant (Tinsley 1975:

---

[39] See (Kragh 1996 pp 318-388) for a review.

[40] In 1970, Robert Dicke demonstrated mathematically that any deviations from flat geometry in the early universe would quickly escalate into a runaway open or closed universe, neither of which is observed (Dicke 1970 p62; Dicke and Peebles 1979).

Gunn and Tinsley 1975; Tinsley 1978). However, uncertainties in galaxy luminosity prevented a clear diagnosis at this point.

8.3 *The theory of inflation*

By the end of the 1970s, studies of the cosmic microwave background indicated a universe that is extremely homogeneous on the largest scales. Yet calculations of astronomical distance indicated that the most distant regions of the universe were simply too far apart to have been in thermal contact, a puzzle known as the *horizon problem*. With the success of gauge theory and spontaneous symmetry breaking in particle physics, several theorists began to consider symmetry-breaking phase transitions in a cosmological context. Thus in the theory of cosmic inflation, it was hypothesized that the infant universe underwent a phase transition that caused an extremely rapid expansion during the first fractions of a second (Guth 1981; Linde 1982; Albrecht and Steinhardt 1982). The hypothesis of inflation had a major impact on theoretical cosmology, as it offered a simple solution to the horizon problem above. In addition, it predicted a universe of flat spatial geometry, offering a solution to the flatness problem identified by Dicke (see section 8.1). Indeed, it was soon shown that inflationary models predicted a spectrum of scale-invariant density perturbations in the early universe consistent with that expected from theories of galaxy formation (Hawking 1982; Starobinksy 1982; Guth and Pi 1982; Bardeen, Steinhardt and Turner 1983).[41]

In the inflationary scenario, the infant universe fleetingly occupies a metastable state known as a false vacuum. The large energy density associated with this state gives rise to an enormous repulsive force that causes the universe to expand as an exponential function of time, a phenomenon that is described by a scalar field that takes the form of an effective cosmological constant. Thus, almost all models of inflation employ a de Sitter spacetime metric, exactly as in the case of steady-state cosmology. In addition, inflation made a startling prediction concerning the cosmological constant in today's universe. Since the model proposed a universe inflated to flat spatial geometry ($\Omega = 1$), while astronomical observations suggested an energy density contribution from matter not more than $\Omega_M = 0.3$, an energy density contribution of about $\Omega_\Lambda = 0.7$ was predicted for the cosmological constant. This facet of the theory of inflation was quickly noticed; indeed, from the mid-1980s onwards, a number of analysts suggested that an inflationary universe of flat geometry, low matter density and positive cosmological constant gave a better fit to astronomical data than the standard Einstein-de Sitter model (Turner, Steigman and Krauss 1984; Peebles 1984; Fujii and Nishioka 1991). In particular, the model offered a solution to the conundrum that open cosmologies with $\Omega_M < 0.3$ predicted temperature fluctuations in the microwave background in excess of $\Delta T/T \sim 10^{-4}$, far above the limit set by observations (Kofman and Starobinsky 1985). However, direct observational evidence for either flatness or a positive cosmological constant was not to emerge for some years (see below).

---

[41] See (Smeenk 2005) for a historical overview of the theory of inflation.

## 9. The cosmological constant in the 1990s

In the 1990s, hints concerning the possible existence of a non-zero cosmic constant began to accumulate. 1992 saw the first reports of the detection of anisotropies in the cosmic microwave background. This data, supplied by sensitive radiometers on board the Cosmic Microwave Background Explorer (COBE) satellite, indicated temperature fluctuations of the order of $\Delta T/T \sim 10^{-5}$ in the background radiation and a spectrum of inhomogeneities in temperature that was apparently scale-invariant (Smoot et al. 1992). These results were beautifully consistent with the standard cold dark matter (CDM) model of structure formation that had emerged in the 1980s[42] and also imposed important new constraints on cosmic models. In particular, the team noted that their data were consistent with a matter-dominated universe with a present Hubble constant $H_0$ less than 50 kms$^{-1}$Mpc$^{-1}$, or with a flat universe with a larger $H_0$ and an energy density dominated by a positive cosmological constant (Wright et al. 1992). Coupled with new constraints on the density of matter from observations of large scale structure and galaxy clustering, interest began to grow in the latter model (Efstathiou et al. 1990; Loveday et al. 1992; Bahcall and Cen 1992; Kofman et al. 1993).

In 1994, new data from the Hubble Space Telescope (HST) and from ground based telescopes suggested an observational value of 80±17 km s$^{-1}$ Mpc$^{-1}$ for the present Hubble constant (Freedman et al. 1994; Pierce et al. 1994). This figure was clearly problematic for cosmic models with $\Omega_M < 0.3$ and $\Omega_\Lambda = 0$, a modern version of the age paradox that had plagued astronomy for so many years (see section 5). However, the result was consistent with a flat cosmology dominated by a cosmological constant. New considerations of gravitational lensing also led some theorists to reconsider the role of the cosmological constant (Krauss and White 1992; Krauss and Schramm 1993). Putting together the revised Hubble constant, the CMB anisotropy data, models of structure formation and constraints on the matter content of the universe set by primordial nucleosynthesis, a number of theorists began to argue forcefully in the 1990s for what was becoming known as the Λ–CDM model of the cosmos (Carroll, Press and Turner 1992; Krauss and Turner 1995; Ostriker and Steinhardt 1995 As new methods of estimating the age of distant globular clusters began to highlight the age paradox of matter-dominated models, interest in models with a cosmological constant continued to grow (Chaboyer et al. 1996; Turner 1997; Turner and White 1997; Krauss 1998).

---

[42] See (Longair 2006 pp 406-410) for a review of the CDM model of structure formation.

*9.1 Supernova candles and dark energy*

From the above, a picture emerges of a slow dawning of Λ–CDM cosmology as a promising model of the universe during the 1990s, at least among theorists. The hypothesis received a dramatic boost at the very end of the century from a new generation of observational programmes to measure the de-acceleration parameter $q_0$. It had been realised for some time that a particularly homogeneous class of supernovae known as SN Ia could serve as ideal standard candles for the measurement of the distance to far flung galaxies and thus offer reliable estimates of the Hubble constant in the distant past. In the late 1980s, a collaboration known as the Supernova Cosmology Project (SCP) was initiated at the Lawrence Berkeley Laboratory in California with the specific aim of probing the time evolution of the Hubble constant using type SN Ia supernovae. After some calibration problems and difficulties in identifying sufficient numbers of supernovae candidates, the team reported promising results[43] and in 1994, a second collaboration known as the High-Z Supernova Search Team (HZT) embarked on a similar program. In 1998 and 1999, both teams reported a result that came as a great surprise to many in the astronomical community; the supernova studies were indicative of a *negative* de-acceleration parameter, i.e., of an acceleration in expansion (Riess et al. 1998; Perlmutter et al. 1999).[44]

The definitive result from the SCP and HZT teams is shown in figure 4. The slight upwards curvature of the redshift/distance relation can be seen in the upper plot; the deviation from linearity can be seen more clearly in the lower plot. These data point unambiguously to an acceleration in expansion over the last five billion years, a phenomenon that was soon dubbed *dark energy* (Turner 1999a, 199b). The cause of this acceleration will be discussed in section 11; for the present we note that the data constitute strong evidence of an acceleration in the rate of cosmic expansion over the last billion years, independent of theoretical models (Shapiro and Turner 2006). In the context of relativistic cosmology, we recall from section 8 that a measurement of $q_0$ is a measure of $\Omega_M - \Omega_\Lambda$; the data from each team suggested a figure of about -0.4 for this quantity. With the matter contribution estimated at $\Omega_M \sim 0.3$, these results were strongly suggestive of an energy contribution from the cosmological constant of the order of $\Omega_\Lambda \sim 0.7$. However, one cosmological parameter remained outstanding – a direct measurement of the spatial geometry of the cosmos.

---

[43] See (Kragh and Overduin 2014 pp 101-105) for a review of early supernova studies.
[44] A description of the reaction of the team-members themselves can be found in (Kirshner 2002 pp 214-224).

## 10. The cosmological constant in the 21st century; the concordance model

The dawn of the 21$^{st}$ century saw yet another important milestone in observational cosmology. In the year 2000, the BOOMERanG (Balloon Observations Of Millimetric Extragalactic Radiation and Geophysics) collaboration, a study of the cosmic microwave background (CMB) using balloon-borne instruments, reported a startling find (de Bernardis et al. 2000); the angular power spectrum of the CMB indicated a cosmos of flat spatial geometry (figure 5).[45] A similar result was reported by the MAXIMA (Millimeter Anisotropy Experiment Imaging Array) collaboration (Balbi et al. 2000; Hanany et al. 2000). The combined data suggested a value of $\Omega \sim 1.0+/-0.04$ for the spatial geometry of the cosmos, strongly indicative of Euclidean geometry (Jaffe et al. 2001).

These results were an important milestone in modern cosmology; assuming a value of $\Omega_M \sim 0.3$ from astronomy, the data pointed directly to a positive cosmological constant with an energy density contribution of the order of $\Omega_\Lambda \sim 0.7$. More quantitatively, the data fitted perfectly with the results from the supernova probes and the Hubble Space Telescope (Jaffe et al. 2001). Put together, the evidence was growing that we inhabit an accelerating cosmos of flat, spatial geometry with energy contributions of $\Omega_M \sim 0.3$ and $\Omega_\Lambda \sim 0.7$ from matter and from the cosmic constant respectively. This result was soon confirmed by satellite observation of the cosmic microwave background, i.e., by the first reports of the Wilkinson Microwave Anisotropy Probe (WMAP) in 2003 (Spergel et al. 2003).

---

[45] As the temperature variations in the CMB represent inhomogeneities in density produced by acoustic waves moving through the cosmic plasma before the epoch of recombination, a measurement of the angular size of the largest compressed sound wave gives a measure of the geometry of the universe (Ferreira 2006 pp 253-254; Longair 2006 pp 415-427).

*10.1 The concordance model*

Since the early years of the 21st century, the results above have been tested by a great number of astronomical experiments such as the Sloan Digital Sky Survey, the Hubble Space Telescope and the Chandra X-ray Observatory. In addition, the WMAP and PLANCK satellite missions have given extremely precise measurements of the cosmic microwave background, the use of gravitational lensing has given important information on galaxy clustering and structure formation, and ever larger supernova studies have given increasingly precise measurements of the time-evolution of the Hubble constant (see (Huterer and Shafer 2017) for a review). These observations have not introduced any substantial changes to the Λ-CDM model outlined above, but have reduced the uncertainties associated with each cosmic parameter. Together, theorists and observers have combined ever more precise measurements of spatial geometry, the rate of cosmic expansion, and the power spectrum of the CMB to give a single model of the cosmos, known as the concordance model. A recent measurement of the power spectrum of the cosmic microwave background by the PLANCK satellite is reproduced in figure 6. As stated in the accompanying paper: *"The temperature and polarization power spectra are consistent with the standard spatially-flat 6-parameter ΛCDM cosmology with a power-law spectrum of adiabatic scalar perturbations. From the Planck temperature data combined with Planck lensing, for this cosmology we find a Hubble constant, $H_0 = (67.8 \pm 0.9)$ km $s^{-1}Mpc^{-1}$, a matter density parameter $\Omega_M = 0.308 \pm 0.012$, and a tilted scalar spectral index with $n_s = 0.968 \pm 0.006$, consistent with the 2013 analysis ... The spatial curvature of our Universe is found to be very close to zero, with $|\Omega_K| < 0.005$...... Combining Planck data with other astrophysical data, including Type Ia supernovae, the equation of state of dark energy is constrained to $w = -1.006 \pm 0.045$, consistent with the expected value for a cosmological constant…The standard big bang nucleosynthesis predictions for the helium and deuterium abundances for the best-fit Planck base ΛCDM cosmology are in excellent agreement with observations"* (Planck Collaboration XIII 2016).

*10.2 The evolution of the concordance model*

From the historian's viewpoint, it is interesting to note that the common narrative[46] of a sudden upheaval in modern cosmology triggered by the discovery of cosmic acceleration in 1998 does not fit well with our findings above. Instead, we have seen that the postulate of a non-zero cosmological constant existed from the very beginning of relativistic cosmology, and its re-emergence as a key component in modern cosmology occurred as the result of a number of observations, from the long search for a possible de-acceleration of cosmic expansion to direct measurements of the geometry of the universe. In addition, the postulate fitted with well-known problems identified in the 1980s and 1990s concerning the age of the universe, the formation of galaxies and the dynamics of galaxy clusters, and with constraints set by considerations of primordial nucleosynthesis. Thus, while the detection of an acceleration in cosmic expansion using supernovae as standard candles undoubtedly marked a watershed in observational cosmology, it is important to note that the 'paradigm shift' to a cosmology with a non-zero cosmological constant occurred as the outcome of a number of advances in both theoretical and observational work.[47]

## 11. The problem of interpretation

While the concordance *Λ-CDM* model represents a great triumph of observational cosmology, the model poses a number of major puzzles for theorists. Three outstanding problems are the nature of the inflationary field (or alternatives), the nature of dark matter and the nature of dark energy. In particular, the empirical evidence for dark energy, mathematically equivalent to a small but non-zero cosmological constant, raises once again the question of the physics underlying this term. Indeed, the so-called *cosmological constant problem* (see section 7) is more pressing than ever as theorists are challenged to explain not only why the cosmological constant is extremely small, but why it has the specific non-zero value that it does; this is sometimes referred to as the *new cosmological constant problem*. A second puzzle is the strange coincidence that the energy contribution of the cosmological constant happens to be of the same order of magnitude as that of matter in today's universe, an enigma that has become known as the *coincidence problem*.

We recall first that calculations by Zel'dovich of the zero-point energy of the quantum vacuum suggested a lower bound of the order of $\rho_\Lambda = 10^{17}$ g/cm$^3$, vastly exceeding the constraints set by observation. This problem was revisited by particle theorists in the 1970s and 80s in the context of gauge theory and spontaneous symmetry breaking. It soon emerged that the problem had got worse, as one could expect additional contributions to the energy density of the vacuum from the phase transitions associated with electro-weak symmetry breaking (Linde 1974; Veltman 1975). Indeed, by the 1980s it had become clear that one could expect an effective cosmological constant $\Lambda_{eff}$ given by

$$\Lambda_{eff} = \Lambda_0 + \Lambda_{zf} + \Lambda_{ew} + \Lambda_{qcd} \qquad (31)$$

where $\Lambda_0$ is a 'bare' (non-quantum) term, $\Lambda_{zf}$ represents the zero point energy of the vacuum, $\Lambda_{ew}$ represents a contribution from the electro-weak phase transition and $\Lambda_{qcd}$ represents a contribution from the quantum chromodynamic phase transition (Hawking 1983, 1984). Of course, one possibility was that the 'bare' cosmological constant could somehow cancel (or nearly cancel) the quantum contributions. However, such a cancellation process seemed exceedingly unlikely as calculations suggested a very large value for each of the quantum terms. Indeed, the American theorist Steven Weinberg found that, under very general conditions, such an approach could not account for a small or vanishing cosmological constant without an extremely high degree of fine tuning, a result that became known as 'Weinberg's no-go theorem' (Weinberg 1989).[48]

---

[46] See for example (Goldsmith 2000 pp 2-6; Kirshner 2002 pp 235-251).

[47] A similar point has recently been made in (Calder and Lahav 2010; Turner and Huterer 2007; Kragh and Overduin 2014 pp 106-107).

One exciting suggestion was that the modern theory of supersymmetry could offer a natural cancellation process that would reduce the quantum energy of the vacuum to zero (Zumino 1975; Hawking 1983). As the theory of supersymmetry proposes the existence of complementary particles of opposite spin to all the known particles, the contributions of the known particles to the energy of the vacuum might be cancelled by that of their super-partners. However, evidence for supersymmetric particles in particle collider experiments has not been forthcoming to date, indicating a symmetry that is broken at energies below the TeV scale (if it exists at all). Since calculations suggest that broken supersymmetry falls far short of reducing Λ to estimates comparable with observation, this approach has fallen from favour (Weinberg 1989; Carroll 2001).

A second approach is the postulate of a scalar field that undergoes a slow decay over time. This possibility goes back to early considerations of the cosmological constant (Gliner 1966; Zel'dovich 1968).[49] In the 1980s, it was noted that such a time-varying field might provide an intriguing link between today's expansion and the hypothesis of inflation in the early universe (Freese et al. 1987; Peebles and Ratra 1988; Ratra and Peebles 1988; Fujii and Nishioka 1991). Interest in such models, which became known as *quintessence*, increased with the emergence of the first hints of empirical evidence for a non-zero cosmological constant (Steinhardt 1997; Caldwell et al. 1998). Following the discovery of unequivocal evidence for cosmic acceleration, a great many candidates for quintessence were proposed, involving dynamical fields from quantum gravity, supergravity and superstring theory (see (Fujii 2000; Steinhardt 2003; Brax 2017) for a review). However, quintessence models are characterized by an equation of state $w$ that varies in time, in contrast with the simple prediction of $w = -1$ for the cosmological constant (see section 7). Thus the hypothesis is tightly constrained by studies of the cosmic microwave background and by considerations of nucleosynthesis in the early universe. In particular, observations of the spectrum of the cosmic microwave background to date do not suggest any variation in the cosmological constant over time, to an accuracy of 0.5 % (Lahav and Liddle 2016; Huterer and Shafer 2017).

---

[48] See (Rugh and Zinkernagel 2002) or (Carroll 2001) for further details.

A third alternative approach to the cosmological constant problem arises from considerations of quantum cosmology, i.e., the study of the universe as a whole as a quantum system. With the use of a wave-function for the universe and Feynman's path-integral approach to quantum mechanics, theorists such as Stephen Hawking, Sidney Coleman and Thomas Banks have found that a distribution of effective values can be derived for the cosmological constant, peaking at $\lambda_{eff} = 0$. (In the language of quantum theory, universes with $\lambda_{eff} = 0$ dominate the path integral, making it probable that the cosmological constant vanishes, or is extremely small). However, this approach involves characterising the cosmological constant as a free parameter, an approach that has been found to be problematic, despite an intriguing proposal by Coleman (Coleman 1988) that the size of the cosmological constant could be set by the interaction of our universe with other universes via wormholes.[50]

The *coincidence problem*, i.e., the relative size of dark energy and the current energy density of matter, presents a second puzzle. As pointed out in section 5, the mean density of matter in the cosmos decreases (in proportion to $R^{-3}$) as the volume expands over time, while the cosmological constant remains effectively unchanged. Thus, it is more than a little surprising that today's value of $\Omega_M$ happens to be of the same order of magnitude as $\Omega_\Lambda$; it seems a cosmic coincidence that begs explanation (Carroll 2001; Peebles and Ratra 2003). Indeed, it could be argued (Steinhardt 2003) that this puzzle is really a modern, generalized version of the flatness problem identified by Dicke in the 1970s (section 8). One exciting explanation could be that dark energy and dark matter are not in fact independent, but interrelated in some subtle way. For example, in the case of some quintessence models, it is proposed that an interaction of the (unknown) quintessence field with matter could yield a natural explanation for the similarity of the energy contributions from matter and dark energy. This idea is reminiscent of that suggested by Bronstein in the 1930s (section 5.2); while many scholars have explored the concept, the results have not been convincing to date (Caldwell et al. 1998; Steinhardt 2003; Kragh and Overduin 2014 pp 82-85).

---

[49] Indeed, the notion of a time-varying cosmic constant was first mooted by Schrödinger (see section 3).
[50] See (Weinberg 1989; Carroll, Press and Turner 1992) for a review of this work.

*11.1 Alternate cosmologies*

Explanations for a small but non-zero cosmological constant have also been offered in the context of alternatives to 'big bang' cosmologies. For example, in the *cyclic universe*, it is proposed that the universe undergoes an endless sequence of cycles from bang to crunch. First considered by early relativists such as Alexander Friedman and Richard Tolman in the context of a universe of closed geometry (Friedman 1922; Tolman 1931a, 1932b), cyclic cosmologies have recently been revived in the context of string theory. In particular, problems concerning infinities in temperature and density and growing entropy are avoided in a scenario where the big bang event is described as a collision of two three-dimensional branes. An exciting aspect of such models is that dark energy is not an ad-hoc addition, but the engine that drives the evolution of the universe throughout its cycles, i.e., a single 'dark energy' scalar field of varying potential is predicted to cause a period of slow accelerated expansion, followed by a period of deceleration and contraction, resulting in a bounce and new cycle (Steinhardt and Turok 2002, 2003). It's worth noting that the cyclic universe offers an alternative to inflation, as each cycle of contraction produces the homogeneity, flatness and energy needed for the next cycle. In addition, the theory provides a simple explanation for the size of the cosmological constant; because the universe is exponentially older than the traditional model, the cosmological constant has had an exponentially long time to decay from the very large value expected from standard calculations of vacuum energy (Steinhardt and Turok 2006). However, cyclic cosmologies predict an equation of state for dark energy that varies in time; such behaviour has not been observable to date as noted above.

Another alternative is the notion of the *emergent universe*. In this scenario, the universe inflates from a static Einstein state after an indefinite period of time, thus reviving the Eddington-Lemaître model (figure 2) in the context of the modern theory of cosmic inflation.[51] This scenario again offers the prospect of an arbitrary long time for the cosmological constant to decay from a very large value (Guendelman 2011; Guendelman and Labraña 2013). However, it once again predicts a time-varying equation of state for dark energy, a phenomenon that has not been observed to date.

*11.2 Philosophical considerations*

We note first that the existence of a small, but non-zero cosmological constant has implications not only for cosmology, but for the general theory of relativity and its description of gravitational phenomena. For example, with $\Lambda \neq 0$ the theory does not reduce to the Minkowski spacetime of special relativity in the absence of a gravitational source (Ray 1990). More generally, the parameter space of solutions of the general field equations is augmented in the case of a non-zero cosmological term, an alteration that may have implications for the description of relativistic phenomena such as gravitational waves and black holes (Ray 1990; Ashtekar 2017).[52] In addition, the term has played an important role in attempts to construct more general field theories, as discussed in section 5.

As regards the physical interpretation of the term, one intriguing philosophical approach to both the 'old cosmological constant' and 'coincidence' problems is to invoke an argument known as the Anthropic Cosmological Principle. In its simplest form, this line of reasoning notes any successful model of our universe can be expected to predict values for cosmic parameters that are compatible with the existence of life within it (Carter 1974; Ellis 2011).[53] For example, Fred Hoyle successfully predicted the existence of a hitherto unsuspected energy state in the carbon atom, on the basis that stellar evolution (and thus life) could not have occurred without it (Barrow and Tipler 1986 pp 252-253; Hoyle 1994 p256).[54]

---

[51] While the Einstein universe is not stable against small perturbations in density, a different scenario may apply where quantized gravitational effects are significant. See (Ellis and Maartens 2004) for an introduction to emergent cosmology.

[52] It has been tacitly assumed that $\Lambda = 0$ in most of the literature describing such phenomena.

[53] Strictly speaking, this type of reasoning is known as the Weak Anthropic Principle (Barrow and Tipler 1986 p16).

[54] It could be argued that this is a rather loose and post-hoc example of anthropic reasoning (Kragh 2010; Ellis 2011).

Applying a similar reasoning to cosmology, Steven Weinberg noted that a value of the cosmological constant corresponding to an energy density more than two orders of magnitude larger than the energy density of matter would not have allowed the formation of the galaxies (Weinberg 1987, 1989). Interest in such reasoning has increased in recent years due to the startling prediction by some models of inflation that the observable universe may comprise only one of a great ensemble of universes (known as the *multiverse*), each with its own set of characteristic cosmic parameters (Vilenkin 1983; Linde 1986, 2008 pp 1-53). In this scenario, a range of values may exist for cosmic parameters and it is thus no surprise that at least one universe exists that is compatible with the existence of observers (Vilenkin 1995; Livio and Rees 2005). However, many physicists take the view that anthropic reasoning can certainly provide useful *constraints* for physical models, but hardly constitutes a true *explanation* of the values of parameters (Pagels 1985; Earman 1987; Kolb and Turner 1990 p269, p315; Deltete 1993). In addition, the principle does not provide specific predictions for the values of particular parameters; for example, one cannot deduce the masses of the elementary particles from anthropic reasoning.[55]

A related philosophical approach to the problem is the assumption that no explanation is required. In this view, the predicted size of the quantum energy of the vacuum is a mathematical shortcoming of quantum field theory, not cosmology, and future advances in quantum theory may reduce the expectation value of the energy of the vacuum to zero. Thus, the observed value of the cosmological constant may be a universal cosmic parameter that is not derivable from theory, i.e., the general theory of relativity contains two constants of nature, the gravitational constant *G* and lambda, both of which are parameters to be determined by measurements (Bludman and Ruderman 1977). After all, the history of science offers many examples where great significance was ascribed to certain numbers that were later found to be an accident of nature. For example, the number of known planets in the solar system was a key parameter in the cosmology of the great astronomer Johannes Kepler; today we view this number of planets as the random outcome of many physical processes. Similarly, little importance is attached today to the so-called Large Number Coincidence identified by Paul Dirac (Dirac 1937). Could it be that the size of the cosmological constant is simply the random outcome of the laws of nature – an *incidence* rather than a *coincidence*? (McCrea 1971; Barrow 2012 pp 292-293; Smeenk 2013). This proposal raises an old question, namely whether the zero-point energy of the quantum vacuum is a real phenomenon or an artefact of theoretical formalism (section 4.1).[56] For example, in recent years, alternative explanations for phenomena such as the Casimir effect have been offered that do not reference the vacuum (Rugh, Zinkernagel and Cao 1999; Jaffe 2005). Turning to cosmology, the conflict between the small size of the cosmological constant observed by astronomers and the predictions of quantum field theory might then be interpreted as strong evidence that the energy associated with the cosmic vacuum is not a real effect (Rugh and Zinkernagel 2002), an argument that is reminiscent of that of Jordan and Pauli (see section 4.1).

---

[55] See (Ellis 2011) or (Smeenk 2013 pp 607-641) for further discussion of anthropic reasoning in cosmology.

*11.3 Quantum gravity*

The most exciting explanation of the cosmological constant problem is that it is a signal of physics beyond the general theory of relativity. This viewpoint was first articulated by the British theorist William McCrea, who suggested that a gravitational theory that did not predict a value for a non-zero cosmological constant (should it turn out to be such) could not be complete (McCrea 1971). Today, theorists have long struggled to incorporate basic tenets of quantum physics into general relativity, or to include the relativistic view of gravity in a unified framework of all the interactions. Indeed, monumental efforts over the last fifty years to bring general relativity and quantum field theory into a single unified framework ('quantum gravity') have been largely unsuccessful, indicating that our theory of gravity is incomplete at least on some scales. In this context, is it surprising that calculations of the quantum energy of the vacuum do not fit seamlessly into relativistic cosmology? Thus, the dramatic conflict between astronomical observations of an acceleration in cosmic expansion (which are essentially model-independent) and theoretical estimates of the zero-point energy of the vacuum (calculated from quantum field theory) may stem from a fundamental clash between the two greatest theories of modern physics - and a true understanding of the physics underlying the cosmological constant will emerge only in the context of quantum gravity (Weinberg 1989; Carroll, Press and Turner 1992; Brax 2017).

*11.4 Future tests*

As the quantum energy of the vacuum is predicted to act in a way that is precisely equivalent to a cosmological constant term, any evidence of a variation over time in the equation of state $p = -\rho c^2$ (or $w = -1$) for dark energy would effectively rule out this candidate. Conversely, observation of a lack of variation in *w* over time would rule out a great many quintessence models. With this in mind, a number of observational programs have been designed to carefully study the behaviour of dark energy over time. In particular, the Dark Energy Survey (DES), operating since 2015, is designed to measure the equation of state *w* using complementary measurements such as galaxy cluster counts, gravitational lensing, galaxy distributions and measurements of cosmological distance using thousands of type Ia supernovae (see (Lahav and Liddle 2016) for further details). Similarly, the Euclid mission of the European Space Agency and Nasa's Joint Dark Energy Mission (JDEM) will probe the time dependence of dark energy using a number of complementary techniques (see (Huterer and Shafer 2017) for a review). Information on dark energy may also arise from experiments in elementary particle physics. For example, any evidence of supersymmetric particles at the Large Hadron Collider could help shed light on the puzzle of the size of the vacuum energy, although the lack of evidence to date for such particles appears to have ruled out many of the simplest supersymmetric models (see (Buckley et al. 2017) for a review).

---

[56] See (Enz 1974) or (Milonni 1994) for a review.

Finally, the recent advent of gravitational-wave astronomy offers an important new test for the modified theories of gravity described above. As many scalar-tensor theories of gravity predict a speed of propagation for gravitational waves that differs from that of light, it was quickly realised that the detection of a gravitational-wave and electromagnetic signal from a single event could in principle break the degeneracy between a cosmological constant and many modified theories of gravity (Lombriser and Taylor 2016; Lombriser and Lima 2017). At the time of writing, this prediction has come to pass. The observation by the gravitational-wave detectors LIGO and VIRGO of a gravitational-wave signal arising from the merger of two neutron stars (Abbott et al. 2017a), and the detection by the Fermi Gamma-ray Space Telescope of an electromagnetic signal arising from the same event (Goldstein et al. 2017) offers strong evidence that gravitational waves propagate at the speed of light to an accuracy of 1x10$^{-15}$ (Abbott et al. 2017b), an observation that imposes severe constraints on theories of modified gravity (Baker et al. 2017; Creminelli and Vernizzi 2017; Ezquiaga and Zumalacárregui 2017; Sakstein and Bhuvnesh 2017).

## 12. Conclusions

The 100-year history of the cosmological constant reveals a fascinating interplay between theoretical physics and astronomical observation. First introduced in order to describe a universe that was assumed to be static, the term was sidelined in simple models of the universe following the discovery of cosmic expansion. However, the term was resurrected on many occasions in the ensuing decades in order to address specific observational puzzles such as the timespan of cosmic expansion, the formation of large-scale structure and the redshifts of quasars; it also found a theoretical motivation in steady-state cosmology and later in the context of the hypothesis of cosmic inflation. In recent years, detailed studies of cosmic expansion, of fluctuations in the cosmic microwave background, of gravitational lensing and of galaxy clustering have presented strong evidence of a universe with an inflationary phase in early times and an accelerated expansion at late times, indicative of a positive cosmological constant dominant at both epochs. Thus we find that the return of the cosmological constant to the forefront of modern cosmology did not occur as an abrupt paradigm shift, but as the result of a number of advances in theory and observation.

However, an understanding of the fundamental physics underlying the cosmological constant remains elusive. Detailed analyses of the simplest explanation, the energy of the vacuum, continue to suggest an estimate for $\Lambda$ that is in dramatic conflict with observation. This result may indicate a major shortcoming in modern field theory, or may be a signal that the general theory of relativity is incomplete. One hundred years after Einstein introduced the term and feared that it would be perceived as a "*mischievous and superfluous stunt*", the cosmological constant poses one of the greatest challenges of modern physics.

**Acknowledgements**

The authors wish to acknowledge the use of online materials provided by the Einstein Papers Project, an important historical resource published by Princeton University Press in conjunction with the California Institute of Technology and the Hebrew University of Jerusalem. We also thank the Hebrew University of Jerusalem for permission to display Einstein's 1917 letter shown in figure 1. Figure 4 is shown by kind permission of the SCP and HZT collaborations, and figures 5 and 6 are shown by permission of the BOOMERanG and Planck collaborations respectively. Cormac O'Raifeartaigh thanks the Dublin Institute for Advanced Studies for the use of research facilities and Norbert Straumann, Pedro Ferreira, Peter Coles and Joe Tenn for helpful discussions. Simon Mitton thanks St Edmund's College, University of Cambridge for the support of his research in the history of science.

**Figure 1.** Excerpt from a letter written by Albert Einstein to Felix Klein on March 26[th], 1917 (Doc. [14- 421] Albert Einstein Archive, ©The Hebrew University of Jerusalem). The second paragraph reads: *"Die neue Variante der Theorie bedeutet formal eine Komplizierung der Grundlagen und wird wohl von fast allen Fachgenossen als ein wenn auch interessantes, aber doch mutwilliges und überflüssiges Kunststück angesehen werden, zumal eine empirische Stütze sich in absehbarer Zeit kaum wird herbeischaffen lassen"* or *"The new version of the theory means, formally, a complication of the foundations and will probably be looked upon by almost all our colleagues as an interesting, though mischievous and superfluous stunt, particularly since it is unlikely that empirical support will be obtainable in the foreseeable future"* (Transl. CPAE 8: Doc. 319).

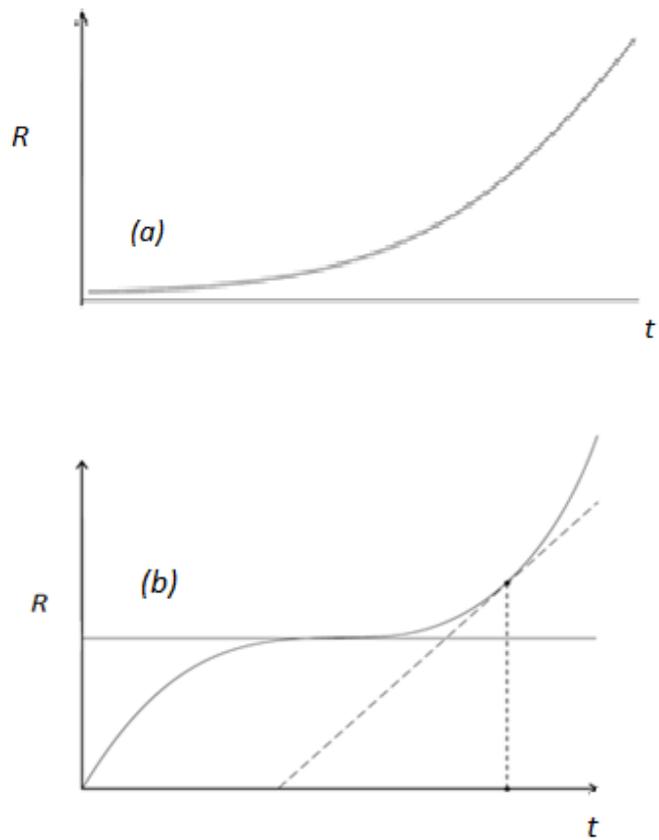

**Figure 2.** Schematic diagrams of **(a)** the Eddington-Lemaître model and **(b)** the Lemaître model. Adapted from (Bondi 1952 p84).

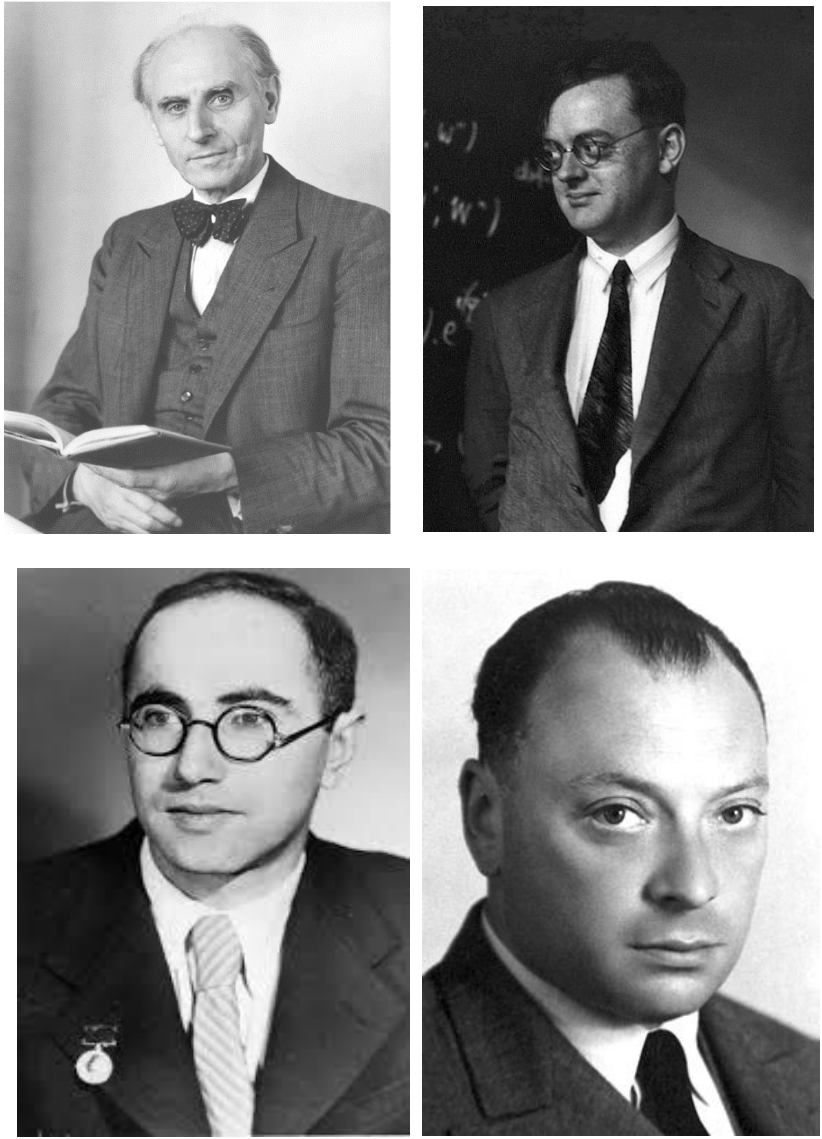

**Figure 3.** Some of the early pioneers who noted that estimates from quantum theory of the energy of the vacuum presented a puzzle for cosmology. Clockwise from top left; Wilhelm Lenz: Pascual Jordan: Wolfgang Pauli (© AIP Emilio Segrè Visual Archives):Yakov Zel'dovic (© CERN photolab).

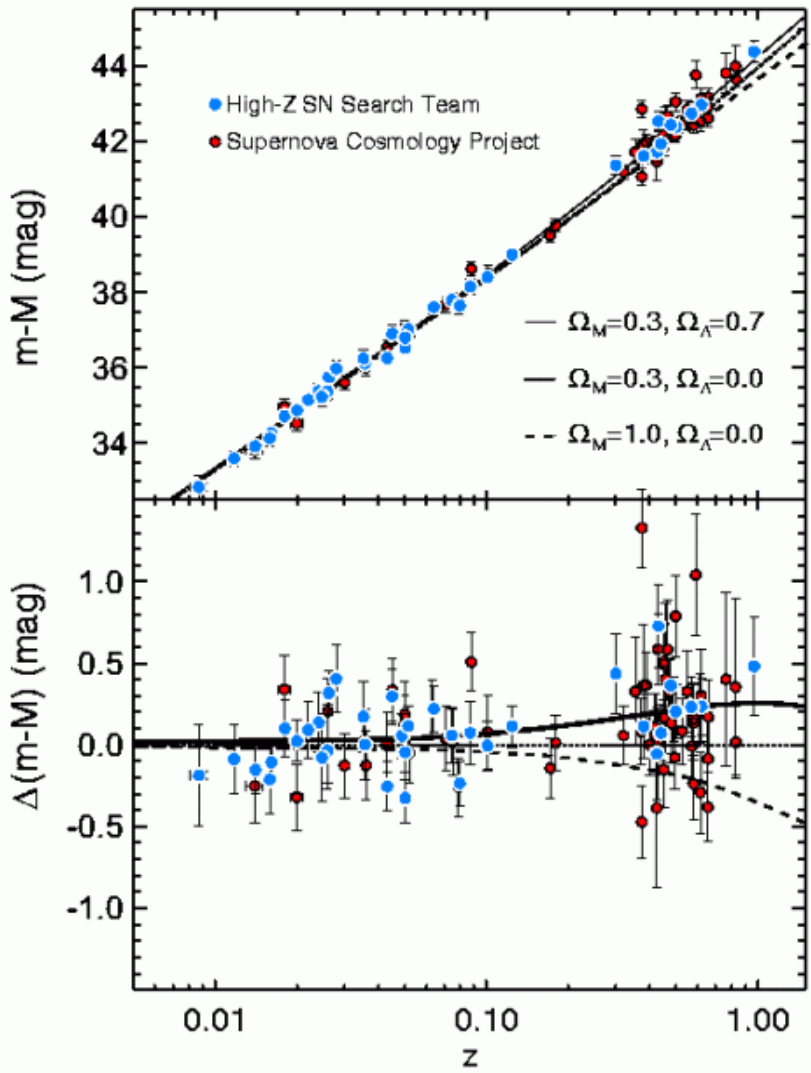

**Figure 4**. Magnitude-redshift plot of Type SN Ia supernovae with data points from the HZT and SCP collaborations. Reproduced from https://www.cfa.harvard.edu/supernova/home.html with permission from the HZT collaboration.

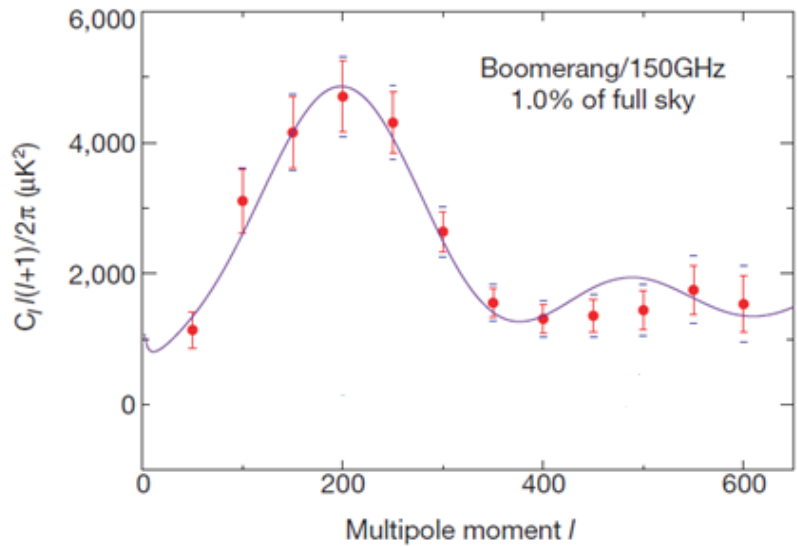

**Fig 5.** Angular power spectrum of the cosmic microwave background measured by the BOOMERanG experiment. The location of the first peak provides strong evidence of a cosmic spatial geometry close to flatness. Adapted from de (Bernardis et al. 2000) with permission from Macmillan Publishers Ltd.

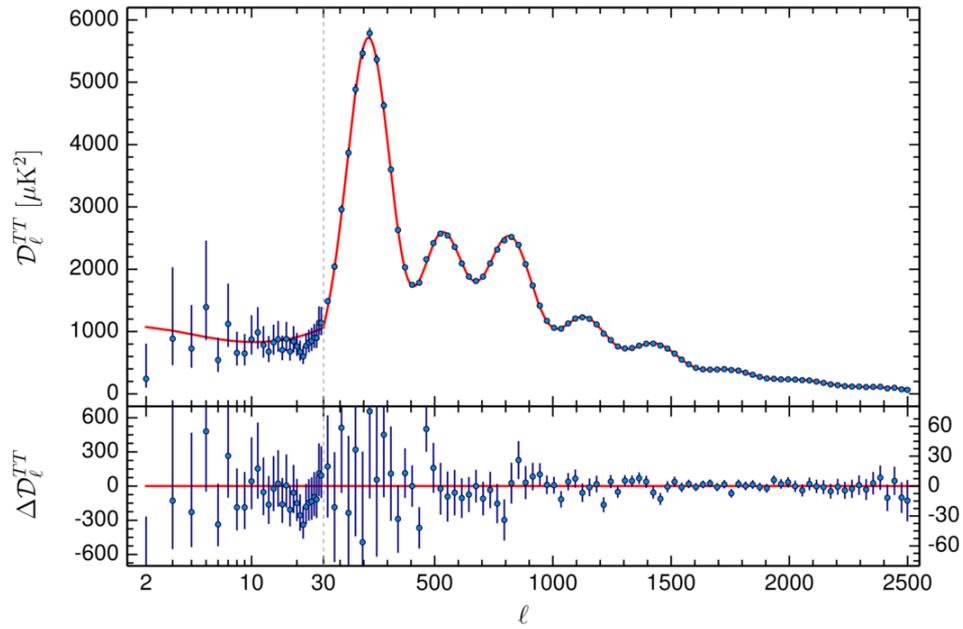

**Figure 6.** Angular power spectrum of the cosmic microwave background (blue dots) measured by the PLANCK satellite (Planck Collaboration XIII 2016). The red line is a thoeretical fit using a Λ-CDM model with the parameters stated in section 10. Residuals with respect to the model are shown in the lower panel.

**References**

Abbott, B.P. et al. 2017a. GW170817: Observation of gravitational waves from a binary neutron star inspiral. *Phys. Rev. Lett*. **119**(16): 161101

Abbott, B.P et al. 2017b. Gravitational waves and gamma-rays from a binary neutron star merger: GW170817 and GRB 170817A. *Astrophys. J. Lett*. **848**(2): L13-18

Abbott, L. 1988. The mystery of the cosmological constant. *Sci. Am*. **258**: 106-113

Albrecht, A. and P.J. Steinhardt. 1982. Cosmology for grand unified theories with radiatively induced symmetry breaking. *Phys. Rev. Lett*. **48**(17): 1220-1223

Alpher, R.A. and R.C. Herman. 1948. On the relative abundance of the elements. *Phys. Rev*. **74**: 1737-1742

Alpher, R.A. and R.C. Herman. 1950. Theory of the Origin and Relative Abundance Distribution of the Elements. *Rev. Mod. Phys*. **22**: 153-212

Alpher, R.A. and R.C. Herman. 1951. Neutron-Capture Theory of Element Formation in an Expanding Universe. *Phys. Rev*. **84**: 60-68

Alpher, R. A., Bethe, H. and G. Gamow 1948. The origin of chemical elements. *Phys. Rev.* **73**(7): 803-804

AP 1931a. Associated Press Report. Prof. Einstein begins his work at Mt. Wilson. *New York Times*, Jan 3, p1

AP 1931b. Associated Press Report. Red shift of nebulae a puzzle, says Einstein. *New York Times*, Feb 12, p2

Ashtekar, A. 2017. Implications of a positive cosmological constant for general relativity. *Rep. Prog. Phys*. **80**(10): 102901-102910

Baade, W. 1952. Extragalactic nebulae. *Trs. IAU* **8**: 397-399

Bahcall, N. A. and R. Cen. 1992. Galaxy clusters and cold dark matter - a low-density unbiased universe? *Astrophys. J.* **398**(2): L81-L84

Baker, T., Bellini, E., Ferreira, P. G., Lagos, M., Noller, J. and I. Sawicki. 2017. Strong constraints on cosmological gravity from GW170817 and GRB 170817A. ArXiv preprint 1710.06394

Balbi et al. 2000. Constraints on Cosmological Parameters from MAXIMA-1. *Astrophys. J.* **545**(1): L1-L4

Bardeen, J. M., Steinhardt and M.S. Turner. 1983. Spontaneous creation of almost scale-free density perturbations in an inflationary universe. *Phys. Rev. D* **28**(4): 679-693

Barrow, J.D. 2012. *The Book of Universes*. Vintage Books, London

Barrow, J.D. and F.J. Tipler. 1986. *The Anthropic Cosmological Principle*. Oxford University Press, Oxford

Belenkiy, A. 2012. Alexander Friedmann and the origins of modern cosmology. *Physics Today* **65** (10): 38-43

Belenkiy, A. 2013. The waters I am entering no one yet has crossed: Alexander Friedmann and the origins of modern cosmology. In *Proceedings of the Conference 'Origins of the Expanding Universe'*. (Eds M. Way and D. Hunter) ASP Conf. Ser. **471:** 71-96

Bergmann, P. 1942. *Introduction to the Theory of Relativity*. Prentice-Hall, New Jersey

Bondi, H. 1952. *Cosmology.* Cambridge University Press, Cambridge

Bondi, H. and T. Gold. 1948. The steady-state theory of the expanding universe. *MNRAS* **108**: 252-70

Bludman, S. A. and M.A. Ruderman. 1977. Induced cosmological constant expected above the phase transition restoring the broken symmetry. *Phys. Rev. Lett*. **38**(5): 255-257

Brax, P. 2017. What makes the universe accelerate? A review on what dark energy could be and how to test it. To be published in *Rep. Prog. Phys*.

Bronstein, M. 1933. On the expanding universe. *Phys. Zeit. Sow*. **3**: 73-82

Buckley, M. R., Feld, D., Macaluso, S., Monteux, A. and D. Shih. 2017. Cornering natural SUSY at LHC Run II and beyond. *JHEP* **2017**(8): 115. ArXiv preprint [1610.08059](1610.08059)

Burbidge, G. R. and E. M. Burbidge 1967. Absorption lines in quasi-stellar objects. *Nature* **216**(5120): 1092-1093

Calder, L. and O. Lahav. 2008. Dark energy: back to Newton? *Astron. Geophys* **49:** 1.13-1.18

Calder, L. and O. Lahav. 2010. Dark energy: how the paradigm shifted. *Physics World* **23**(1): 32-37

Caldwell, R. R., Dave, R. and P. J. Steinhardt 1998. Cosmological imprint of an energy component with general equation of state. *Phys. Rev. Lett.* **80**(8): 1582-1585

Carroll, S. M. 2001. The cosmological constant. *Liv. Rev. Rel*. **4**: 1-56

Carroll S. M., Press, W. H. and E. L. Turner 1992. The cosmological constant. *Ann. Rev. Astron. Astrophys.* **30**: 499-542

Carter, B. 1974. Large number coincidences and the anthropic principle in cosmology. In *Confrontation of Cosmological Theories with Observational Data; Proceedings of the 1973 IAU Symposium* (Ed. M.S. Longair) Reidel, Dordrecht pp 291-298. Republished in *Gen. Rel. Grav.* **43**(11): 3225-323 (2011)

Carter, B. 2006. Anthropic principle in cosmology. In *Current Issues in Cosmology* (Eds J-C. Pecker and J. Narlikar) Cambridge University Press, Cambridge pp 173 - 179

Casimir, H.B.G. 1948. On the attraction between two perfectly conducting plates. *Proc. K. Ned. Akad. Wet*. **51**(7): 793-780

Chaboyer, B., Demarque, P., Kernan, P. J. and L.M. Krauss 1996. A lower limit on the age of the universe. *Science* **271**(5251): 957-961.

Coleman, S. 1988. Why there is nothing rather than something; a theory of the cosmological constant. *Nuc Phys. B* **310**(3): 643-668

Creminelli, P. and F. Vernizzi. 2017. Dark energy after GW170817. ArXiv preprint 1710.05877

de Bernardis, P. et al. 2000. A flat universe from high-resolution maps of the cosmic microwave background radiation. *Nature* **404**(6781): 955–959.

de Sitter, W. 1917. On Einstein's theory of gravitation and its astronomical consequences. Third paper. *MNRAS* **78:** 3-28

de Sitter, W. 1930a. Proceedings of the RAS. *The Observatory* **53**: 37-39

de Sitter, W. 1930b. On the distances and radial velocities of the extragalactic nebulae, and the explanation of the latter by the relativity theory of inertia. *PNAS.* **16:** 474-488

de Sitter, W. 1930c. The expanding universe. Discussion of Lemaître's solution of the equations of the inertial field. *Bull. Astron. Inst. Neth*. **5** (193): 211-218

de Sitter, W. 1931. The expanding universe. *Scientia* **49**: 1-10

de Sitter, W, 1932. *Kosmos: A Course of Six Lectures on the Development of Our Insight into the Structure of the Universe*. Harvard University Press, Cambridge MA

Deltete, R. J. 1993. What does the anthropic principle explain? *Persp. Sci*. **1**: 285-305

Dicke, R. H. 1970. *Gravitation and the Universe: Jayne Lectures for 1969*. American Philosophical Society.

Dicke, R. H. and P.J.E. Peebles 1979. The big bang cosmology - enigmas and nostrums. In *General Relativity; an Einstein Centenary Survey* (Eds S.W. Hawking and W. Israel) Cambridge University Press pp. 504 - 517

Dirac, P. A. M. 1937. The cosmological constants. *Nature* **139** (3512): 323

Earman, J. 1987. The SAP also rises: a critical examination of the anthropic principle. *Am. Phil. Quart*. **24**(4): 307- 317

Earman, J. 2001. Lambda: the constant that refuses to die. *Arch. Hist. Ex. Sci*. **55**: 189-220

Eddington, A.S.1930. On the instability of Einstein's spherical world. *MNRAS* **90**: 668-678

Eddington A.S. 1931a. The recession of the extra-galactic nebulae. *MNRAS* **92**: 3-6

Eddington, A. S. 1931b. On the value of the cosmical constant. *Proc. Roy. Soc*. **A133**: 605-615

Eddington, A.S. 1933. *The Expanding Universe*. Cambridge University Press, Cambridge

Eddington, A,S. 1944. The recession-constant of the galaxies. *MNRAS* **104**: 200-204

Efstathiou, G., Sutherland W. J. and S. J. Maddox 1990. The cosmolgical constant and cold dark matter. *Nature* **348:** 705-707

Einstein, A. 1915a. Die Feldgleichungen der Gravitation. *Sitz. König. Preuss. Akad*. 844-847. Or 'The field equations of gravitation' CPAE **6** (Doc. 25)

Einstein, A. 1915b. Erklärung der Perihelbewegung des Merkur aus der allgemeinen Relativitätstheorie. *Sitz. König. Preuss. Akad*. 831-839. Or 'Explanation of the perhelion motion of Mercury from the general theory of relativity' CPAE **6** (Doc. 24)

Einstein, A. 1916. Die Grundlage der allgemeinen Relativitätstheorie. *Ann. Physik*. **49:** 769-822. Or 'The foundation of the general theory of relativity' CPAE **6** (Doc. 30)

Einstein, A. 1917a. Kosmologische Betrachtungen zur allgemeinen Relativitätstheorie. *Sitz. König. Preuss. Akad*. 142-152. Or 'Cosmological considerations in the general theory of relativity' CPAE **6** (Doc. 43).

Einstein, A. 1917b. Letter to Willem de Sitter, March 12th. CPAE **8** (Doc. 311).

Einstein, A. 1917c. Letter to Felix Klein, March 26th. CPAE **8** (Doc. 319).

Einstein, A. 1917d. Letter to Willem de Sitter, April 14th. CPAE **8** (Doc. 325).

Einstein, A. 1918a. Bemerkung zu Herrn Schrödingers Notiz "Über ein Lösungssystem der allgemein kovarianten Gravitationsgleichungen". *Phys. Zeit.* **19:** 165-166. Or 'Comment on Schrödinger's Note "On a system of solutions for the generally covariant gravitational field equations" ' CPAE **7** (Doc. 3).

Einstein 1918b. Letter to Michele Besso, July 29th. CPAE **8** (Doc. 591).

Einstein 1918c. Letter to Michele Besso, August 20th. CPAE **8** (Doc. 604).

Einstein 1918d. *Über die Spezielle und die Allgemeine Relativitätstheorie*. Vieweg (Braunschweig). 3rd Edition. CPAE **6** (Doc.42).

Einstein, A. 1918e. Kritisches zu einer von Hrn. De Sitter gegebenen Lösung der Gravitationsgleichungen. *Sitz. König. Preuss. Akad*. 270-272. Or 'Critical comment on a solution of the gravitational field equations given by Mr. de Sitter' CPAE **7** (Doc. 5).

Einstein, A. 1919a. Spielen Gravitationsfelder im Aufbau der materiellen Elementarteilchen eine wesentliche Rolle? *Sitz. König. Preuss. Akad*. 349-356. Or 'Do gravitation fields play an essential part in the structure of the elementary particles of matter?' CPAE **7** (Doc. 17).

Einstein, A. 1919b. Bermerkung über periodischen Schwankungen der Mondlänge, welche bisher nach der Newtonschen Mechanik nicht erklärbar schienen. *Sitz. König. Preuss. Akad*.

433-436. Or 'Comment about periodical fluctuations of lunar longitude, which so far appeared to be inexplicable in Newtonian mechanics' CPAE 7 (Doc. 18).

Einstein, A. 1921a. *Geometrie und Erfahrung*. Springer, Berlin. Or 'Geometry and Experience'. CPAE 7 (Doc. 52).

Einstein, A. 1921b. Eine einfache Anwendung des Newtonschen Gravitationsgesetzes auf die kugelförmigen Sternhaufen. In *Festschrift der Kaiser-Wilhelm-Gesellschaft zur Förderung der Wissenschaften*. Springer, Berlin 50-52. Or 'A simple application of the Newtonian law of gravitation to globular star clusters' CPAE 7 (Doc. 56)

Einstein, A.1922a. *Vier Vorlesungen über Relativitätstheorie*. Vieweg, Berlin. Or *The Meaning of Relativity*. Methuen, London (Transl. E.Adams). CPAE 7 (Doc. 71)

Einstein, A. 1922b. Bemerkung zu der Arbeit von A. Friedmann "Über die Krümmung des Raumes" *Zeit. Phys.* **11**: 326. Or 'Comment on A. Friedmann's paper "On The Curvature of Space" ' CPAE 13 (Doc. 340).

Einstein, A.1923a. Notiz zu der Arbeit von A. Friedmann "Über die Krümmung des Raumes" *Zeit. Phys.* **16**: 228. Or 'Note to the paper by A. Friedmann "On the Curvature of Space" ' CPAE 14 (Doc. 51).

Einstein, A. 1923b. Notiz zu der Arbeit von A. Friedmann "Über die Krümmung des Raumes" . The Albert Einstein Archives. Doc. **1**-26.

Einstein, A. 1923c. Postcard to Hermann Weyl, May 23rd. CPAE 14 (Doc. 40).

Einstein, A. 1931a. Zum kosmologischen Problem der allgemeinen Relativitätstheorie. *Sitz. König. Preuss. Akad*. 235-237. Eng. transl. (O'Raifeartaigh and McCann 2014).

Einstein, A. 1931b. Zum kosmologischen Problem. *Albert Einstein Archive Online*, Doc. [2-112]. http://alberteinstein.info/vufind1/Record/EAR000034354. Eng. transl. (O'Raifeartaigh et al. 2014).

Einstein, A. 1933. Sur la structure cosmologique de l'espace (Fr. transl. M. Solovine). In *'La Théorie de la Relativité'*, Hermann, Paris. (Eng. transl. O'Raifeartaigh et al. 2015).

Einstein, A. 1945. On the *'cosmologic problem'*. Appendix I to *The Meaning of Relativity*. Princeton University Press, Princeton (3rd Ed.) 112-135

Einstein, A. and W. de Sitter. 1932. On the relation between the expansion and the mean density of the universe. *PNAS* **18** (3): 213-214

Ellis, G.F.R. 2003. A historical review of how the cosmological constant has fared in general relativity and cosmology. *Cha. Sol. Fract.* **16**: 505-512

Ellis, G. F. R. 2011. Editorial note to: Brandon Carter, Large number coincidences and the anthropic principle in cosmology. *Gen. Rel. Grav.* **43**(11): 3213-3223

Ellis, G.F.R. and R. Maartens. 2004. The emergent universe: inflationary cosmology with no singularity. *Class. Quant. Gravity* **21**: 223 – 239

Enz, C.P. 1974. Is the Zero-Point Energy Real? In *Physical Reality and Mathematical Description* (Eds C. P. Enz and J. Mehra) Reidel, Dordrecht pp. 124–132

Enz, C.P. and A. Thellung. 1960. Nullpunktsenergie und Anordnung nicht vertauschbarer Faktoren im Hamiltonoperator. *Helv. Phys. Acta* **33**: 839-848

Ezquiaga, J.M. and M. Zumalacárregui. Dark energy after GW170817: Dead ends and the road ahead. *Phys. Rev. Lett*. **119**: 251304-251315

Pedro Ferreira, P. 2007. *The State of the Universe: A Primer in Modern Cosmology*. Phoenix, London pp 251-254

Freedman, W. L. et al. 1994. Distance to the Virgo cluster galaxy M100 from Hubble Space Telescope observations of Cepheids. *Nature* **371**(6500): 757-762

Freese, K., Adams, F. C., Frieman, J. A. and E. Mottola. 1987. Cosmology with decaying vacuum energy *Nuc. Phys. B* **287**: 797-814

Friedman, A. 1922. Über die Krümmung des Raumes. *Zeit. Physik.* **10**: 377-386. Available in English translation as 'On the curvature of space' *Gen. Rel. Grav.* **31**(12): 1991-2000 (1999)

Friedman, A. 1923. *Die Welt als Raum und Zeit*. Deutsch, Berlin. Eng trans. *The World in Space and Time*. Minkowski Press, 2014.

Frieman J.A., Turner, M.S. and D. Huterer. 2008. Dark energy and the accelerating universe. *Ann. Rev. Astron. Astrophys.* **46**(1): 385-432

Fujii, Y. 2000. Quintessence, scalar-tensor theories and non-Newtonian gravity. *Phys. Rev. D* **62**(4): 4011-22. ArXiv preprint [9911064](#)

Fujii, Y. and T. Nishioka. 1991. Reconciling a small density parameter to inflation. *Phys Lett B* **254**: 347-350

Gamow, G. 1942. Concerning the origin of chemical elements. *JWAS* **32**(12): 353-35

Gamow, G. 1946. Expanding universe and the origin of elements. *Phys. Rev*. **70**(7-8): 572-573

Gamow, G. 1949. On relativistic cosmogony. *Rev. Mod. Phys*. **21**(3): 367-373

Gamow, G. 1956. The evolutionary universe. *Sci. Am.* **195**(3): 136-156

Gamow, G. 1970. *My World Line: An Informal Autobiography*. Viking Press, New York

Gamow, G. and E. Teller. 1939a. On the origin of great nebulae. *Phys Rev* **55**: 654-657

Gamow, G. and E. Teller. 1939b. The expanding universe and the origin of the great nebulæ. *Nature* **143** (3612): 116-117

Goldsmith, D. 2000. *The Runaway Universe: the Race to Find the Future of the Cosmos*. Basic Books, New York

Goldstein, A. et al. 2017. An ordinary short gamma-ray burst with extraordinary implications: Fermi-GBM detection of GRB 170817A. *Astrophys. J. Lett*. **848**(2): L14-28

Gott, J.R., Gunn J.E., Schramm, D.N. and B.M. Tinsley. 1974. An unbound universe. *Astrophy. J*. **194:** 543-553

Gliner, E.B. 1966. Algebraic properties of the energy-momentum tensor and vacuum-like states of matter. *JETP* **22:** 378-382

Guendelman, E. I. 2011. Non-singular origin of the universe and the cosmological constant problem. *Int. J. Mod. Phys. D* **20**(14): 2767-2771

Guendelman, E. I. and P. Labraña. 2013. Connecting the nonsingular origin of the universe, the vacuum structure and the cosmological constant problem. *Int. J. Mod. Phys. D* **22**(9): 13300181-133001835

Gunn, J. E. 1975. On the mean mass density in the universe. *Ann. NY Acad. Sci*. **262**: 21-29

Gunn, J. E. and J.B. Oke, J. B. 1975. Spectrophotometry of faint cluster galaxies and the Hubble diagram: an approach to cosmology. *Astrophys. J*. **195**: 255-268

Gunn, J.E. and B.M. Tinsley 1975. An accelerating universe. *Nature* **257**: 454-457

Guth, A. H. 1981.The inflationary universe: a possible solution for the horizon and flatness problems. *Phys. Rev. D* **23:** 347-356

Guth, A. H. and S.-Y. Pi. 1982. Fluctuations in the new inflationary universe. *Phys. Rev. Lett*. **49**: 1110-1113

Hanany, S. et al. 2000. MAXIMA-1: A measurement of the cosmic microwave background anisotropy on angular scales of 10'-5°. *Astrophys J*., **545**(1): L5-L9

Harvey, A. 2009. Dark energy and the cosmological constant: a brief introduction. *Eur. J. Phys*. **30**: 877–889

Harvey, A. 2012a. The cosmological constant. ArXiv preprint 1211.6337

Harvey, A. 2012b. How Einstein discovered dark energy. ArXiv preprint 1211.6338

Harvey, A. and E. Schucking 2000. Einstein's mistake and the cosmological constant *Am. J. of Phys*. **68**: 723-728

Hawking, S. W. 1983. The cosmological constant. *Phil. Trans. Roy. Soc. Lond*. **A310** (1512): 303-309

Hawking, S.W. 1984. The cosmological constant is probably zero. *Phys Lett. B* **134** (6): 403-404

Heckmann, O. 1931. Über die Metrik des sich ausdehnenden Universums. *Nach. Gesell. Wiss. Göttingen, Math.-Phys. Klasse* **2**: 126-131

Heckmann, O. 1932. Die Ausdehnung der Welt in ihrer Abhängigkeit von der Zeit. *Nach. Gesell. Wiss. Göttingen, Math.-Phys. Klasse* **2**: 181-190

Hoyle, F. 1948. A new model for the expanding universe. *MNRAS* **108**: 372-382

Hoyle, F. 1994. *Home Is Where The Wind Blows: Chapters From A Cosmologists's Life*. University Science Books, California.

Hoyle, F. and Burbidge, G.R. 1966. Relation between the redshifts of quasi-stellar objects and their radio magnitudes. *Nature* **212**: 1334

Hoyle, F. and A. Sandage 1956. The second-order term in the redshift-magnitude relation. *Pub. Ast. Soc. Pac.* **68** (403): 301-307

Hubble, E. 1925. Cepheids in spiral nebulae. *The Observatory* **48**: 139-142

Hubble, E. 1929. A relation between distance and radial velocity among extra-galactic nebulae. *PNAS.* **15**: 168-173

Hubble, E. and M.L. Humason. 1931. The velocity-distance relation among extra-galactic nebulae. *Astrophys. J*. **74**: 43-80

Humason, M. L., Mayall, N. U. and A. R. Sandage. 1956. Redshifts and magnitudes of extragalactic nebulae. *Astron. J.* **61**: 97-162

Huterer, D. and D. L. Shafer. 2017. Dark energy two decades after: observables, probes, consistency tests. To be published in *Rep. Prog. Phys*. ArXiv preprint 1709.01091

Jackson, J. C. 1970. The dynamics of clusters of galaxies in universes with non-zero cosmological constant, and the virial theorem mass discrepancy. *MNRAS* **148**: 249-260

Jaffe, A. H. et al. 2001. Cosmology from MAXIMA-1, BOOMERANG, and COBE DMR cosmic microwave background observations. *Phys. Rev. Lett*. **86**(16): 3475-3479

Jaffe, R. L. 2005. Casimir effect and the quantum vacuum. *Phys. Rev. D* **72**(2): 021301

Jordan, P. and W. Pauli 1928. Zur Quantenelektrodynamik ladungsfreier Felder. *Zeit. Phys*. **47**: 151–173

Kardashev, N. 1967. Lemaître's universe and observations. *Astrophys. J.* **150**: L135-L139

Kazanas, D. 1980. Dynamics of the universe and spontaneous symmetry breaking. *Astrophys. J Lett*. **150** : L135-145

Kirshner, R.P. 2002. *The Extravagant Universe: Exploding Stars, Dark Energy and the Accelerating Cosmos*. Princeton University Press, Princeton.

Kofman, L. A. and A.A. Starobinsky 1985. Effect of the cosmological constant on largescale anisotropies in the microwave background. *Sov. Ast. Lett*. **11**: 271-274.

Kofman, L. Gnedin N. and N. Bahcall. 1993. Cosmological constant, COBE cosmic microwave background anisotropy, and large-scale clustering. *Astrophys. J*. **413**(1): 1-9

Kolb, E.W. and M.S. Turner 1990. *The Early Universe*. Addison-Wesley, New York.

Kragh, H.S. 1996. *Cosmology and Controversy*. Princeton University Press, Princeton

Kragh, H.S. 2007. *Conceptions of Cosmos: From Myths to the Accelerating Universe: A History of Cosmology*. Oxford University Press, Oxford

Kragh, H. 2010. An anthropic myth: Fred Hoyle's carbon-12 resonance level. *Arch. Hist. Ex. Sci.* **64**(3): 721-751

Kragh, H.S. 2012. Preludes to dark energy: zero-point energy and vacuum speculations. *Arch. Hist. Ex. Sci.* **66**(3): 199-240

Kragh, H. 2015. On Arthur Eddington's theory of everything. ArXiv preprint 1510.04046

Kragh, H. and D. Lambert. 2007. The context of discovery: Lemaître and the origin of the primeval-atom universe. *Ann. Sci.* 445-470

Kragh, H. S. and J.M. Overduin. 2014. *The Weight of the Vacuum: A Scientific History of Dark Energy*. Springer, Berlin.

Krauss, L. M. 1998. The end of the age problem, and the case for a cosmological constant revisited. *Astrophys. J.* **501**:461-466

Krauss, L. M. and D. N. Schramm. 1993. Angular diameters as a probe of a cosmological constant and Omega. *Astrophys J.* **405**(2): L43-L46

Krauss L.M. and M.S. Turner. 1995. The cosmological constant is back. *Gen. Rel. Grav*. **27** (11): 1137-1144

Krauss, L. M. and M. White 1992. Gravitational lensing, finite galaxy cores, and the cosmological constant. *Astrophys J.* **394**(2): 385-395

Lahav, O. and A.R. Liddle. 2016. The cosmological parameters 2016. In *The Review of Particle Physics* (Particle Data Group). *Chin. Phys C* **40**(10): 386-393

Laplace, P-S. 1846. *Mécanique Céleste* **5**. Book 16, p.481

Lemaître, G. 1925. Note on de Sitter's universe. *J. Math. Phys*. **4:** 188-192

Lemaître, G. 1927. Un univers homogène de masse constante et de rayon croissant, rendant compte de la vitesse radiale des nébuleuses extra-galactiques. *Ann. Soc. Sci. Brux.* **A47:** 49-59. See also (Luminet 2013)

Lemaître, G. 1931a. A homogeneous universe of constant mass and increasing radius, accounting for the radial velocity of the extra-galactic nebulae. *MNRAS* **91:** 483-490

Lemaître, G. 1931b. The beginning of the world from the point of view of quantum theory. *Nature* **127**: 706

Lemaître, G. 1931c. The expanding universe. *MNRAS* **91:** 490-501

Lemaître, G. 1931d. L'expansion de l'espace. *Rev. Quest. Sci.* **20**: 391-410.

Lemaître, G. 1933. L' universe en expansion. *Ann. Soc. Sci. Brux* **A53:** 51-85. Eng. transl. 'The expanding universe' *Gen. Rel. Grav.* **29**(5): 641-680 (1997)

Lemaître, G. 1934. Evolution of the expanding universe. *PNAS* **20:** 12-17

Lemaître, G. 1949. The cosmological constant. In *Albert Einstein: Philosopher Scientist, The Library of Living Philosophers* **VII** (Ed. P.A. Schilpp). George Banta, Wisconsin pp 439-456

Lemaître, G. 1958. Recontres avec Einstein. *Rev. Quest. Sci.* **129:** 129-132

Lenz, W. 1926. Das Gleichgewicht von Materie und Strahlung in Einsteins geschlossener Welt. *Phys. Zeit.* **27:** 642–645

Linde, A. D., 1974. Is the Lee constant a cosmological constant? *JETP Lett*. **19**: 183-184

Linde, A.D. 1982. A new inflationary universe scenario: a possible solution of the horizon, flatness, homogeneity, isotropy and primordial monopole problems. *Phys. Lett. B* **108**(6): 389- 393

Linde, A.D. 1984. The inflationary universe. *Rep. Prog. Phys*. **47**: 925-986,

Linde, A. D. 1986. Eternal chaotic inflation. *Mod. Phys. Lett. A* **1** (02): 81-85

Linde, A. D. 2008. Inflationary cosmology. In *Lecture Notes in Physics* **738**. Springer, Berlin.

Livio, M. 2013. *Brilliant Blunders: from Darwin to Einstein*. Simon & Schuster, New York

Livio, M. and M. J. Rees. 2005. Anthropic reasoning. *Science* **309**: 1022-1023

Lombriser, L. and A. Taylor. 2016. Breaking a dark degeneracy with gravitational waves. *JCAP* **03.** 031

Lombriser, L. and N. A. Lima. 2017. Challenges to self-acceleration in modified gravity from gravitational waves and large-scale structure. *Phys Lett. B* **765**: 382-385

Longair, M. S. 2006. *The Cosmic Century: A History of Astrophysics and Cosmology*. Cambridge University Press, Cambridge.

Longair, M. S. and P.A.G. Scheuer 1967. Red-shift magnitude relation for quasi-stellar objects. *Nature* **215** (5104): 919-922

Loveday, J., Efstathiou, G., Peterson, B. A. and S. J. Maddox. Large-scale structure in the universe - results from the Stromlo-APM redshift survey. *Astrophys J*. **400**(2): L43-L46.

Luminet, J-P. 2013. Editorial note to 'A homogeneous universe of constant mass and increasing radius, accounting for the radial velocity of the extra-galactic nebulae'. *Gen. Rel. Grav*. **45**(8): 1619-1633

Maneff, G. 1932. Über das kosmologische Problem der Relativitätstheorie. *Zeit. Astrophys*.

**4**: 231–240

McCrea,W.H. 1951. Relativity theory and the creation of matter. *Proc. Roy. Soc.* **A206** (1087): 562–575

McCrea, W. H. 1971. The cosmical constant. *Q. J. Roy. Ast. Soc*. **12:** 140-153

Milne, E. A. 1933. World-Structure and the Expansion of the Universe. *Zeit. Astrophys.* **6**: 1-95

Milne, A. 1935. *Relativity, Gravitation and World Structure*. Clarendon Press, Oxford

Neumann, C. 1896. *Allgemeine Untersuchungen über das Newton'sche Prinzip der Fernwirkungen*. Teubner, Leipzig

Milonni, P.W. 1994. *The Quantum Vacuum*. Academic, New York

North, J.D. 1965. *The Measure of the Universe: A History of Modern Cosmology*. Oxford University Press

Nernst, W. 1916. Über einen Versuch, von quantentheoretischen Betrachtungen zur Annahme stetiger Energieänderungen zurückzukehren. *Verh. Dtsch. Phys. Ges*. **18**: 83–116

Norton, J. D. 1999. The cosmological woes of Newtonian gravitation theory. In *'The Expanding Worlds of General Relativity: Einstein Studies Vol.7'* (Eds H. Goenner et al.) Birkhäuser, Boston pp 271-322

Nussbaumer, H. 2014a. Einstein's conversion from his static to an expanding universe. *Eur. Phys. J (H)* **39**(1): 37-62

Nussbaumer, H. 2014b. Einstein's aborted model of a steady-state universe. To be published in "In memoriam Hilmar W. Duerbeck" *Acta Historica Astronomiae.* (Eds W. Dick et al.). ArXiv preprint 1402.4099

Nussbaumer, H. and L. Bieri. 2009. *Discovering the Expanding Universe*. Cambridge University Press, Cambridge

Oort, J. 1932. The force exerted by the stellar system in a direction perpendicular to the galactic plane and some related problems. *Bull. Astron. Inst. Neth*. **6**: 249-287.

O'Raifeartaigh, C. and B. McCann. 2014. Einstein's cosmic model of 1931 revisited; an analysis and translation of a forgotten model of the universe. *Eur. Phys. J (H)* **39**(1): 63-85

O'Raifeartaigh, C., McCann, B., Nahm, W. and S. Mitton. 2014. Einstein's steady-state theory: an abandoned model of the cosmos. *Eur. Phys. J (H)* **39**(3):353-369.

O'Raifeartaigh, C., O'Keeffe, M., Nahm, W. and S. Mitton. 2015. Einstein's cosmology review of 1933: a new perspective on the Einstein-de Sitter model of the cosmos. *Eur. Phys. J (H)* **40**(3): 301-336

O'Raifeartaigh, C., O'Keeffe, M., Nahm, W. and S. Mitton. 2017. Einstein's 1917 static model of the cosmos: a centennial review. *Eur. Phys. J (H)* **42**(3): 431-474

Ostriker, J.P. and S. Mitton. 2013. *Heart of Darkness: Unravelling the Mysteries of the Invisible Universe.* Princeton University Press, Princeton

Ostriker, J. P. and P.J. Steinhardt. 1995. The observational case for a low-density Universe with a non-zero cosmological constant. *Nature* **377**(6550): 600-602

Pagels, H.R. 1985. A cozy cosmology. *The Sciences* **25**(2): 34-38

Pauli, W. 1933. Die allgemeinen Prinzipien der Wellenmechanik. In *Handbuch der Physik, Quantentheorie* **24**(1) (Eds. H. Bethe et al.) Springer, Berlin pp 83-272

Pauli, W. 1946. Exclusion principle and quantum mechanics. In *Nobel Lectures in Physics 1942-1962*. Elsevier, Amsterdam (1964)

Pauli, W. 1958. *Theory of Relativity*. Pergamon Press, New York

Peebles, P. J. E. 1976. A cosmic virial theorem. *Astrophys. J. Lett*. **205**: L109-113

Peebles, P.J.E. 1984. Tests of cosmological models constrained by inflation. *Astrophys J.* **284**: 439-444

Peebles, P. J. E. 1986. The mean mass density of the universe. *Nature* **321**: 27-32

Peebles, P. J. E. and B. Ratra. 1988. Cosmology with a time-variable cosmological 'constant'. *Astrophys. J. Lett* **325:** L17-L20.

Peebles, P.J.E. and B. Ratra 2003. The cosmological constant and dark energy. *Rev. Mod. Phys.* **75**(2): 559-606

Perlmutter S. et al. 1999. Measurements of Ω and Λ from 42 high redshift supernovae. *Astrophys. J.* **517**: 565-586

Petrosian, V. 1974. Confrontation of Lemaître models and the cosmological constant with observations. In *Confrontation of Cosmological Theories with Observational Data*: *Proceedings of the 1973 IAU Symposium* (Ed. M. Longair) Reidel, Dordrecht.

Petrosian V. and E. Salpeter. 1970. Lemaître models and the cosmological constant. *Comm. Ast. Sp. Phy.* **2**: 109-115

Petrosian V., Salpeter E. and P. Szekeres. 1967. Quasi-stellar objects in universes with non-zero cosmological constant. *Astrophys. J.* **147**: 1222-1226

Pierce, M. J. et al. The Hubble constant and Virgo cluster distance from observations of Cepheid variables. *Nature* **371**(6496): 385-389

Planck, M. 1911. Eine neue Strahlungshypothese. *Verh. Dtsch. Phys. Ges*. **13**: 138–148

Planck Collaboration XIII. 2016. Cosmological parameters. *Astron. & Astrophys*. **594** (A13):1-63

Ratra, B. and P. J. E. Peebles. Cosmological consequences of a rolling homogeneous scalar field. *Phys. Rev. D* **37** (12): 3406-3427

Ray, C. 1990. The cosmological constant: Einstein's greatest mistake? *Stud. Hist. Phil. Sci. A* **21**(4): 589-604

Realdi, M. and G. Peruzzi 2009. Einstein, de Sitter and the beginning of relativistic cosmology in 1917. *Gen. Rel. Grav.* **41** (2): 225-247

Riess, A. G. et al. 1998. Observational evidence from supernovae for an accelerating universe and a cosmological constant. *Astron. J.* **116**: 1009-1038

Rindler, W. 1969. *Essential Relativity: Special, General, and Cosmological*. Van Nostrand, New York

Robertson, H.P. 1932. The expanding universe. *Science* **76**: 221-226

Robertson, H.P. 1933. Relativistic cosmology. *Rev. Mod. Phys*. **5**(1): 62-90

Robertson, H. P. 1935. Kinematics and world-structure. *Astrophys. J.* **82**: 284-301

Robertson, H. P. 1955. The theoretical aspects of the nebular redshift. *Pub. Ast. Soc. Pac.* **67** (395): 82-98

Rowan-Robinson, M. 1968. On cosmological models with an antipole. *MNRAS* **141**: 445-458

Rugh, S. E., Zinkernagel H. and T.Y. Cao. 1999. The Casimir effect and the interpretation of the vacuum. *Stud. Hist. Phil. Mod. Phys*. **30**(1): 111-139

Rugh, S. E. and H. Zinkernagel. 2002. The quantum vacuum and the cosmological constant problem. *Stud. Hist. Phil. Mod. Phys*. **33**(4): 663-705

Sakstein, J. and J. Bhuvnesh. 2017. Implications of the neutron star merger GW170817 for cosmological scalar-tensor theories. ArXiv preprint 1710.05893

Sandage, A. R. 1958. Current problems in the extragalactic distance scale. *Astrophys. J*. **127:** 513-526.

Sandage, A.R. 1961. The ability of the 200-inch telescope to discriminate between selected world models. *Astrophys. J*. **133**: 355-389

Sandage, A. R. 1962. The change of redshift and apparent luminosity of galaxies due to the deceleration of selected expanding universes. *Astrophys. J*. **136**: 319-333

Sandage, A.R. 1965. The existence of a major new constituent of the universe: the quasistellar galaxies. *Astrophys. J*. **141**: 1560-1578

Sandage, A. R. 1970. Cosmology: a search for two numbers. *Physics Today* **23**(2): 34-42

Sandage, A.R. 1995. Practical cosmology: inventing the past. In *The Deep Universe* (Eds Sandage et al.) Springer, Berlin pp 1-232.

Sandage, A.R. and G.A. Tammann 1984. The dynamical parameters of the universe. In *Proceedings of the 1983 ESO/CERN Symposium* (Eds G. Setti and L. Van Hove): 127-147

Schmidt, M. 1963. 3C 273: A star-like object with large red-shift. *Nature* **197** (4872): 1040-1050

Schmidt, M. 1965. Large redshifts of five quasi-stellar sources. *Astrophys. J.* **141**: 1295-1300

Schmidt, M. and T. A. Matthews. 1964. Redshift of the quasi-stellar radio sources 3C 47 and 3C 147. *Astrophys. J.* **139**: 781 -785

Schmidt, B.G. et al. 1998. The high-z supernova search: measuring cosmic deacceleration and global curvature of the universe usng type 1a supernovae. *Astrophys. J.* **507**: 46-63

Schrödinger, E. 1918. Über ein Lösungssystem der allgemein kovarianten Gravitationsgleichungen. *Phys. Zeit.* **19:** 20-22. Transl. excerpts in (Harvey 2012b).

Schulmann R., Kox A.J., Janssen M. and J. Illy. 1998. The Einstein-deSitter-Weyl-Klein debate. In CPAE 8A p351

Seeliger, H. von 1895. Über das Newton'sche Gravitationsgesetz. *Astron. Nach.* **137**: 129-136.

Seeliger, H. von 1896. Über das Newton'sche Gravitationsgesetz *Sitz. König. Bayer. Akad. Wiss.* **126**: 373-400

Seeliger, H. von. 1898a. On Newton's law of gravitation. *Pop. Astron.* **5**: 474-478

Seeliger, H. von. 1898b. On Newton's law of gravitation. *Pop. Astron.* **5**: 544-551

Seitter, W. C. and R. Duemmler. 1989. The cosmological constant - historical annotations. In *Morphological Cosmology; Proceedings of the Eleventh Krakow Cosmological School* (Eds P. Flin and H. Duerbeck) Springer, Berlin pp. 377-387.

Shapiro, C. and M. S. Turner. 2006. What do we really know about cosmic acceleration? *Astrophys. J.* **649** (2): 563-569

Shklovsky, J. 1967. On the nature of "standard" absorption spectrum of the quasi-stellar objects. *Astrophys. J.* **150**: L1-3

Slipher, V. M. 1915. Spectrographic observations of nebulae. *Pop. Ast.* **23**: 21-24

Slipher, V. M. 1917. Nebulae. *Proc. Am. Phil. Soc.* **56**: 403-409

Smeenk, C. 2005. False vacuum: early universe cosmology and the development of inflation. In *The Universe of General Relativity*: *Einstein Studies Vol. 11*. (Eds A. J. Kox and J. Eisenstaedt) Birkhäuser, Boston pp 223-258

Smeenk, C. 2013. Philosophy of Cosmology. In *The Oxford Handbook of Philosophy of Physics*. (Ed. R. Batterman), Oxford University Press, Oxford.

Smeenk, C. 2014. Einstein's role in the creation of relativistic cosmology. In *The Cambridge Companion to Einstein*. (Eds M. Janssen and C. Lehner). Cambridge University Press, Cambridge pp 228-269

Smith, R. 1982. *The Expanding Universe: Astronomy's Great Debate 1900-1931*. Cambridge University Press, Cambridge

Smoot, G. et al. 1992. Structure in the COBE differential microwave radiometer first-year maps. *Astrophys J.* **396**(1): L1-L5.

Sparnaay, M.J. 1957. Attractive forces between flat plates. *Nature* **180**(4581): 334-344

Spergel, D. N. et al. (2003). First-year Wilkinson Microwave Anisotropy Probe (WMAP) observations: determination of cosmological parameters. *Astrophys. J. Suppl.* **148** (1): 175–194

Starobinsky, A.A. 1982. Dynamics of phase transition in the new inflationary universe scenario and generation of perturbations. *Phys. Lett. B* **117**: 175-178

Steinhardt, P. J. 1997. Cosmological challenges for the 21st century. In *Critical Problems in Physics: Proceedings of a Conference Celebrating the 250th Anniversary of Princeton University* (Eds V. L. Fitch et al.) Princeton University Press, Princeton, p.123

Steinhardt, P. J. 2003. A quintessential introduction to dark energy. *Roy. Soc. Lon. Trans.* **A361**(1812): 2497-2513

Steinhardt, P. J. and N. Turok. 2002. A cyclic model of the universe. *Science* **296** (5572): 1436-1439

Steinhardt, P.J. and N. Turok. 2003. The cyclic universe: an informal introduction *Nuclear Physics B Proc. Suppl.* **124**: 38-49

Steinhardt, P. J. and N. Turok. 2006. Why the cosmological constant is small and positive. *Science* **312** (5777): 1180-1183

Straumann, N. 1999. The mystery of the cosmic vacuum energy density and the accelerated expansion of the Universe. *Eur. J. Phys.* **20**(6): 419-427

Straumann, N. 2002. The history of the cosmological constant problem. In *On the Nature of Dark Energy: Proceedings of the 18th IAP Astrophysics Colloquium*. (Ed. P. Brax et al.) Frontier Group (Paris). [ArXiv preprint 0208027](#)

Straumann, N. 2013. *General Relativity*. Springer, Berlin (2$^{nd}$ ed.)

Tammann, G. A. 1979. Precise determination of the distances of galaxies. *IAU Coll.* **54**: 263-293

Tammann, G. A., Sandage, A. R. and A. Yahil.1979. The determination of cosmological parameters. *Lecture Notes for the 1979 Les Houches Summer School*, Basel, Switzerland.

Taylor E.F. and J.A. Wheeler. 2000. *Exploring Black Holes: Introduction to General Relativity.* Addison Wesley, San Francisco.

Tinsley, B. M. 1975. The evolution of galaxies and its significance for cosmology. *Ann. NY Acad. of Sci*. **262**: 436-448

Tinsley, B. M. 1978. Accelerating universe revisited. *Nature* **273**: 208-211

Tolman, R. 1929. On the astronomical implications of the de Sitter line element for the universe. *PNAS* **69:** 245-274

Tolman, R.C. 1930. More complete discussion of the time-dependence of the non-static line element for the universe. *PNAS.* **16:** 409-420

Tolman, R.C. 1931a. On the theoretical requirements for a periodic behaviour of the universe. *Phys. Rev.* **38:** 1758-1771

Tolman, R.C. 1931b. Letter to Albert Einstein. September 14[th]. Albert Einstein Archive. 23-31

Tolman, R. C. 1932. Models of the Physical Universe. *Science* **75** (1945): 367-373

Tolman, R.C. 1934. *Relativity, Thermodynamics and Cosmology*. Oxford University Press, Oxford.

Tolman, R.C. and M. Ward 1932. On the behaviour of non-static models of the universe when the cosmological term is omitted. *Phys. Rev.* **39:** 835-843

Topper, D.R. 2013. *How Einstein Created Relativity out of Physics and Astronomy*. Springer, New York.

Turner, M.S. 1997. The Case for ΛCDM. ArXiv preprint [9703161](9703161)

Turner, M.S. 1999a. Cosmology solved? Maybe. *Nuc. Phys. B Proc. Suppl*. **72**(1-3): 69-80

Turner, M.S. 1999b. Dark Matter and dark energy in the Universe. In *The Third Stromlo Symposium: The Galactic Halo* (Eds. B.K. Gibson et al.) ASP Conf. Ser. **165**: 431-435

Turner, M. S. and D. Huterer. 2007. Cosmic acceleration, dark energy, and fundamental physics. *J. Phys. Soc. Jap*. **76**(11): 10151-10159

Turner, M. S. and M. White. 1997. CDM models with a smooth component. *Phys. Rev. D* **56** (8): 4439-4443

Turner, M.S., Steigman G. and L.M. Krauss. 1984. Flatness of the universe: reconciling theoretical prejudices with observational data. *Phys. Rev. Lett*. **52**(23): 2090-2093

Veltman, M., 1975. Cosmology and the Higgs mass. *Phys. Rev. Lett*. **34**(12): 777-779

Vilenkin, A. 1983. Birth of inflationary universes. *Phys. Rev. D* **27** (12): 2848-2855

Vilenkin, A. 1995. Predictions from quantum cosmology. *Phys. Rev. Lett*. **74**(6): 846-849

Walker, A.G. 1937. On Milne's world structure. *Proc. Lond. Math. Soc*. **S242**(1): 90-12

Weinberg, S. 1972. *Gravitation and Cosmology: Principles and Applications of the General Theory of Relativity*. Wiley & Sons, New York

Weinberg, S. 1987. Anthropic bound on the cosmological constant. *Phys. Rev. Lett*. **59**(22): 2607-2610

Weinberg, S. 1989. The cosmological constant problem. *Rev. Mod. Phys.* **61**(1): 1-23

Weyl, H. 1918. Gravitation und Elektrizität. *Sitz. König. Preuss. Akad.*: 465-478

Wright et al. 1992. Interpretation of the cosmic microwave background radiation anisotropy detected by the COBE Differential Microwave Radiometer. *Astrophys. J*. **396**(1): L13-L18.

Zaycoff, R. 1932: Zur relativistichen Kosmogonie. *Zeit. Astrophys*. **6:** 128–197

Zel'dovich Y. B. 1967. Cosmological constant and elementary particles. *JETP Lett.* **6**: 316-317

Zel'dovich Y. B. 1968. The cosmological constant and the theory of elementary particles. *Sov. Phys. Usp.* **11**: 381-393. Republished with editorial notes in *Gen. Rel. Grav*. **40:** 1557–1591 (2008)

Zwicky, F. 1933. Die Rotverschiebung von extragalaktischen Nebeln. *Helv. Phys. Acta*. **6**: 110-127

Zwicky, F. 1937. On the masses of nebulae and of clusters of nebulae. *Astrophys. J*. **86**: 217-246

Zumino, B. 1975. Supersymmetry and the vacuum. *Nucl. Phys B* **89** (3): 535-546